\documentclass[prd,twocolumn,showpacs,preprintnumbers,amsmath,amssymb]{revtex4}
\usepackage{graphicx}
\usepackage{epsfig}
\usepackage{rotating}
\usepackage{dcolumn}
\usepackage{bm}

\newcommand{\PRE}[1]{}       

\newcommand{\fb}{\, {\rm fb}}
\newcommand{\lsim}
{{\;\raise0.3ex\hbox{$<$\kern-0.75em\raise-1.1ex\hbox{$\sim$}}\;}}
\newcommand{\gsim}
{{\;\raise0.3ex\hbox{$>$\kern-0.75em\raise-1.1ex\hbox{$\sim$}}\;}}
\newcommand{\tev}{\, {\rm TeV}}
\newcommand{\gev}{\, {\rm GeV}}
\def\lesq{\ell_e^2}
\def\resq{r_e^2}
\def\lefour{\ell_e^4}
\def\refour{r_e^4}

\def\lqsq{\ell_f^2}
\def\rqsq{r_f^2}
\def\pdot#1#2{s_{#1#2}}
\def\btilZ{\widetilde b_Z}
\def\btilW{\widetilde b_W}
\def\reaZbZconj{\Re(a_Z \, b_Z^*)}
\def\imaZbZconj{\Im(a_Z \, b_Z^*)}
\def\reaZbtilZconj{\Re(a_Z \,\btilZ^*)}
\def\imaZbtilZconj{\Im(a_Z \, \btilZ^*)}
\def\reaWbWconj{\Re(a_W \, b_W^*)}
\def\imaWbWconj{\Im(a_W \, b_W^*)}
\def\reaWbtilWconj{\Re(a_W \,\btilW^*)}
\def\imaWbtilWconj{\Im(a_W \, \btilW^*)}

\begin{document}
\preprint{IISc-CHEP/10/05}
\preprint{hep-ph/0509070}
\title{\PRE{\vspace*{2.0in}}
Signatures of anomalous $VVH$ interactions at a linear collider}
\PRE{\vspace*{.4in}
}

\author{Sudhansu S. Biswal$^{1,}$\footnote{Electronic address: sudhansu@cts.iisc.ernet.in}}
\author{Debajyoti Choudhury$^{2,}$\footnote{Electronic address: debchou@physics.du.ac.in}}
\author{Rohini M. Godbole$^{1,}$\footnote{Electronic address: rohini@cts.iisc.ernet.in}}
\author{Ritesh K. Singh$^{1,}$\footnote{Electronic address: ritesh@cts.iisc.ernet.in}\PRE{\vspace{.3in}}
}

\vspace*{.3in}
\affiliation{{}$^1$Center for High Energy Physics, Indian Institute 
  of Science, Bangalore, 560012, India\\ \break 
$^2$Department of Physics and Astrophysics, 
    University of Delhi, 
  Delhi 110007, India\break and \break
  HarishChandra Research Institute, Chhatnag Road, Jhusi, 
  Allahabad 211019, India}
\begin{abstract}
  We examine, in a model independent way, the sensitivity of a Linear Collider
  to the couplings of a light Higgs boson to gauge bosons.  Including the
  possibility of $CP$ violation, we construct several observables that probe
  the different anomalous couplings possible.  For an intermediate mass Higgs,
  a collider operating at a center of mass energy of 500 GeV and with an
  integrated luminosity of 500 fb$^{-1}$ is shown to be able to constrain the 
  $ZZH$ vertex at the few per cent level, and with even higher sensitivity 
  in certain directions. However, the lack of sufficient number of observables
  as well as contamination from the $ZZH$ vertex limits the precision with 
  which the $WWH$ coupling can be measured.
\end{abstract}
\pacs{14.80.Cp, 14.70.FM, 14.70.Hp}
\maketitle
\noindent
\section{Introduction}
Although the standard model (SM) has withstood all  possible experimental
challenges and has been tested to an unprecedented degree of accuracy, so far
there has been no direct experimental verification of the phenomenon of
spontaneous symmetry breaking. With the latter being considered a central
pillar of this theory and its various extensions, the search for a Higgs boson
is one of the main aims for many current and future
colliders~\cite{Godbole:2002mt}.  Within the SM, the only fundamental spin-$0$
object is the ($CP$-even) Higgs boson and remains the only particle in the SM
spectrum to be found yet. Rather, a lower bound on the mass of the SM
Higgs boson, (about 114.5 GeV) is provided by the direct searches at the LEP
collider~\cite{higmin}.
Electroweak precision measurements, on the other hand, provide an upper bound
on its mass of about 204 GeV at 95\% C.L.~\cite{higmax}. It should be realized
that both these limits are model dependent and may be relaxed in extensions of 
the SM. For example, the lower limit can be relaxed in generic 2-Higgs doublet
models~\cite{2hdm} or 
in models with $CP$ violation~\cite{cp-mssm}.
In the latter case, direct searches at LEP and elsewhere still allow the
lightest Higgs boson to be as light as 10 GeV~\cite{lowerH}.
Similarly, the upper bound on
the mass of the (lightest) Higgs in some extensions may be substantially
higher~\cite{highH}.  The Large Hadron
Collider (LHC) is expected to be capable~\cite{lhc-tdr} of searching for 
the Higgs boson in the entire mass range allowed.

It is then quite obvious that just the discovery of the Higgs boson at the LHC
will not be sufficient to validate the minimal SM. For one, the only neutral
scalar in the SM is a $J^{CP} = 0^{++}$ state arising from a $SU(2)_L$ doublet
with hypercharge 1, while its various extensions can have several Higgs bosons
with different $CP$ properties and $U(1)$ quantum numbers.  The minimal
supersymmetric standard model (MSSM), for example, has two $CP$-even states and 
a single $CP$-odd one~\cite{rmg}. Thus, should a neutral spin-0 state be observed
at the LHC,
a study of its $CP$-property would be essential to establish it as {\it the} SM
Higgs boson~\cite{godbole}.

Since, at an $e^+e^-$ collider, the dominant production modes of a
neutral Higgs boson proceed via its coupling with a pair of gauge bosons
($VV, \ V=W,Z$), any change in the $VVH$ couplings from their 
SM values can be probed via such production processes. 
Within the SM/MSSM, the only (renormalizable) interaction term involving the
Higgs boson and a pair of gauge bosons is the one arising from the Higgs
kinetic term. However, once we accept the SM to be only an effective
low-energy description, higher-dimensional (and hence non-renormalizable)
terms are allowed. If we only demand Lorentz invariance and gauge invariance,
the most general coupling structure may be expressed as
\begin{eqnarray}
\Gamma_{\mu\nu} &=& g_V\left[a_V \ g_{\mu\nu}+\frac{b_V}{m_V^2}(k_{1\nu} k_{2\mu}
 - g_{\mu\nu}  \ k_{1} \cdot k_{2}) \right.\nonumber\\
&&\left.+\frac{\tilde{b_V}}{m_V^2} \ \epsilon_{\mu\nu\alpha\beta} \ 
k_1^{\alpha}  k_2^{\beta}\right]
\label{eq:coup}
\end{eqnarray}
where $k_i$ denote the momenta of the two $W$'s ($Z$'s), $
g_W^{SM}~=~e \ \cot\theta_W M_Z $ and $g_Z^{SM}~=~2 \ e \
M_Z/\sin2\theta_W$.  In the context of the SM, at the tree level, $
a_W^{SM}~=a_Z^{SM}~=~1$ while the other couplings vanish
identically. At the one-loop level or in a different theory, effective
or otherwise, these may assume significantly different values. We
study this most general set of anomalous couplings of the Higgs boson to a
pair of $W$s and $Z$ at a linear collider (LC) in the processes
$e^+e^-\to f \bar f H$, with $f$ being a light fermion.

The various kinematical distributions for the process $e^+e^-\to f \bar f H$,
proceeding via vector boson fusion and Higgsstrahlung, with unpolarized 
beams has
been studied in the context of the SM~\cite{dist}. The effect of beam
polarization has also been investigated for the SM~\cite{pol}.  The anomalous
$ZZH$ couplings have been studied in Refs.\cite{rattazzi,stong,He:2002qi,
Hagiwara:1993ck,Gounaris:1995mx} for the LC and in 
Refs.\cite{Zhang:2003it,plehn}
for the LHC in terms of higher dimensional operators. 
Ref.~\cite{Hagiwara:2000tk} investigates the possibility to probe the
anomalous $VZH$ couplings, $V=\gamma,Z$, using the optimal observable
technique~\cite{oot} for both polarized and un-polarized beams. 
Ref.~\cite{t.han}, on the other hand, probes the $CP$-violating coupling
$\tilde b_Z$ by means of asymmetries in kinematical distributions and beam
polarization. In Ref.~\cite{maria}, the $VVH$ vertex is studied in the process
of $\gamma\gamma \to H\to W^+W^-/ZZ$ using angular distributions of the decay 
products.

The rest of the paper is organized as follows. In section~\ref{sec:vvh} we
discuss the possible sources and symmetries of the anomalous $VVH$
couplings and the rates of various processes involving these couplings.  In
section~\ref{sec:zzh} we construct
several observables with appropriate $CP$ and $\tilde T$ property to
probe various $ZZH$ anomalous couplings. In section~\ref{sec:wwh} we
construct similar observables to probe anomalous $WWH$ couplings,
which  we then use along with the ones constructed for the $ZZH$
case. Section~\ref{sec:res} contains a discussion and summary of our findings.

%
%
%
\section{The $VVH$ couplings}
     \label{sec:vvh}
The anomalous $VVH$ couplings in Eq.(\ref{eq:coup}) can appear from various
sources such as via higher order corrections to the vertex in a
renormalizable theory~\cite{kniehl:NPB} or from higher dimensional operators 
in an effective theory~\cite{operator}.
For example, in the MSSM, the non-zero phases of the trilinear SUSY breaking
parameter $A$ and the gaugino/higgsino mass parameters can induce
$CP$-violating terms in the scalar potential at one loop level even
though the tree level potential is $CP$-conserving. As a consequence,
the Higgs-boson mass eigenstates can turn out to be linear
combinations of $CP$-even and -odd states. This modifies the effective
coupling of the Higgs boson to the known particles from what is
predicted in the SM (or even from that within a version of MSSM with
no $CP$-violation accruing from the scalar sector).

\begin{figure}
\epsfig{file=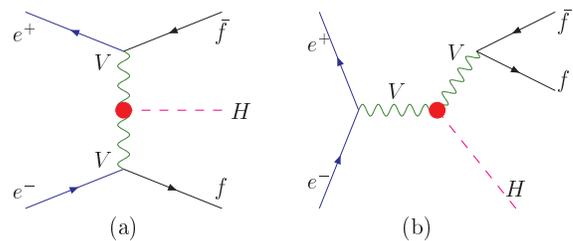,width=8.0cm}
\caption{\label{fig:feyn}Feynman diagrams for the process $e^+e^- \to f\bar f
H$; (a) is $t$-channel or fusion diagram, while (b) is $s$-channel
or Bjorken diagram. For $f=e,\nu_e$ both (a) \& (b) contribute while for the 
all other fermions only (b) contributes.}
\end{figure}

In a generic multi-doublet model, whether 
supersymmetric~\cite{sumrule} or otherwise~\cite{Choudhury:2003ut}, 
the couplings of the neutral Higgs bosons to 
a pair of gauge bosons satisfy the sum rule
$$\sum_i a^2_{VVH_i} = 1.$$ Thus, while $a_{VVH_i}$ for a given Higgs
boson can be significantly smaller than the SM value, any violation of
the above sum rule would indicate either the presence of higher
$SU(2)_L$ multiplets or more complicated symmetry breaking structures
(such as those within higher-dimensional
theories)~\cite{Choudhury:2003ut}.  The couplings
$b_V$ or $\tilde b_V$ can arise only at a higher order in a
renormalizable theory~\cite{kniehl:NPB}.  Furthermore, within models
such as the SM/MSSM where the tree-level scalar potential is
$CP$-conserving, $\tilde b_V$ may be generated only at an order of
perturbation theory higher than that in which the Higgs sector
acquires $CP$-violating terms.  However, in an effective theory, which
satisfies $SU(2) \otimes U(1)$ symmetry, the couplings $b_V$ and
$\tilde b_V$ can arise, at the lowest order, from terms such as
$F_{\mu \nu} \, F^{\mu \nu} \, \Phi^\dagger \Phi$ or $F_{\mu \nu} \,
\tilde F^{\mu \nu} \, \Phi^\dagger \Phi$~\cite{operator} where $\Phi$
is the usual Higgs doublet, $F_{\mu \nu}$ is a field strength tensor
and $\tilde F^{\mu \nu}$ its dual.  It can be easily ascertained that
the effects of the higher dimensional terms in the trilinear vertices
of interest can be absorbed into $b_V$ ($\tilde b_V$) by ascribing
them with non-trivial momentum-dependences (form factor behavior).
Clearly, if the cut-off scale $\Lambda$ of this theory is much larger
than the typical energy at which a scattering experiment is to be
performed, the said dependence would be weak.  In all processes that
we shall be considering, this turns out to be the case. In particular,
the Bjorken process (Fig.~\ref{fig:feyn}(b)) essentially proceeds at a
fixed center-of-mass energy, hence both $b_Z$ and $\tilde b_Z$ are
constant for this process. Even for the other processes of interest,
namely gauge boson fusion (Fig.~\ref{fig:feyn}(a)), the momentum
dependence of the form-factors have a rather minor role to play,
especially for $\Lambda \gsim 1 \tev$. This suggests that we can treat
$a_V, b_V, \tilde b_V$ as phenomenological and energy-independent
parameters.
\subsection{Symmetries of anomalous couplings}
A consequence of imposing an $SU(2) \otimes U(1)$ symmetry would be to
relate the anomalous couplings, $b_W$ and $\tilde b_W$, for the $WWH$ vertex 
with those for the $ZZH$ vertex. However, rather than attempting to calculate
these couplings within a given model, we shall treat them as purely 
phenomenological inputs, whose effect on the kinematics of various
final states in collider processes can be analyzed. 
In general, each of these couplings can be complex, reflecting
possible absorptive parts of the loops, either from the SM or from
some new high scale physics beyond the SM. It is easy to see that a
non-vanishing value for either $\Im(b_V)$ or $\Im(\tilde b_V)$
destroys the hermiticity of the effective theory. Such couplings can
be envisaged when one goes beyond the Born approximation, whence they
arise from final state interactions, or, in other words, out of the
absorptive part(s) of higher order diagrams, presumably mediated by
new physics. A fallout of non-hermitian transition matrices is
non-zero expectation values of observables which are odd under
$CP\tilde{T}$, where $\tilde{T}$ stands for the pseudo-time reversal
transformation, one which reverses particle momenta and spins but does
not interchange initial and final states. Of course,
such non-zero expectation values will be indicative of final state
interaction only when kinematic cuts are such that the phase space
integration respects $CP\tilde{T}$.  Note that $a_V$ too can be
complex in general and can give an additional $\tilde T$-odd
contribution.  However, for the processes that we will consider, the
phase of at least one of $a_W$ and $a_Z$ can always be rotated away,
and we make this choice for $a_Z$.  Henceforth, we shall assume that
$a_W$ and $a_Z$ are {\em close to their SM value}, i.e.  $a_i =
1+\Delta a_i$, the rationale being that any departure from $a_W^{SM}$
and $a_Z^{SM}$ respectively would be the easiest to measure

Unfortunately, this still leaves us with many free parameters making
an analysis cumbersome.  One might argue that $SU(2) \otimes U(1)$
gauge invariance would predict $\Delta a_W = \Delta a_Z$.  
However, once symmetry breaking effects are considered, this does not
necessarily follow~\cite{kniehl:NPB}. Nevertheless, we will make this 
simplifying assumption that $\Delta a_W$ is real and equal to $\Delta a_Z$,
i.e. $a_W=a_Z$, since the equality is found to hold true in some specific 
cases~\cite{awaz} (and would be dictated if $SU(2) \otimes U(1)$ were 
to be an exact symmetry of the effective theory). With this assumption, we
list, in Table~\ref{tab:coup}, 
the $CP$ and $\tilde{T}$ properties of such operators.  

\begin{table}[t]
\caption{\label{tab:coup}Transformation properties of various anomalous
couplings under discrete transformations.}
\begin{ruledtabular}
\begin{tabular}{cccccc}
Trans. & $a_V$  & $\Re(b_V)$ & $\Im(b_V)$ &
$\Re(\tilde b_V)$ & $\Im(\tilde b_V)$\\ \hline
$CP$ & $+$  & $+$ & $+$& $-$ & $-$ \\ 
$\tilde T$ & $+$  & $+$ & $-$& $-$ & $+$ 
\end{tabular}
\end{ruledtabular}
   \label{table:prop}
\end{table}

Finally, keeping in view
the higher-dimensional nature of all of these couplings, we retain
only contributions up to the lowest non-trivial order, arising from
terms linear in the additional couplings.
\subsection{Cross-sections}
The dominant channels of Higgs production at an electron-positron colliders
are
 \begin{enumerate}
 \item the 2-body Bjorken process ($e^+e^-\to ZH$);
 \item in association with a pair of neutrinos ($e^+e^-\to \nu_e\bar\nu_e H$),
       {\em i.e.} $W$-fusion;
 \item in association with an $e^+e^-$ pair ($e^+e^-\to e^+e^- H$), {\em i.e.}
       $Z$-fusion
 \end{enumerate}
Note that the Bjorken-process also contributes to the other two final states
through the subsequent decay of the $Z$. 
Of these three, the $e^+e^-H$ channel is considerably suppressed (by almost a
factor of 10) with respect to the $\nu_e\bar\nu_e H$ channel over a very
wide range of center-of-mass energies ($\sqrt{s}$) and Higgs masses.
As can be expected, at large $\sqrt{s}$, the Bjorken process suffers the
usual $s$-channel suppression and has a smaller cross section compared
to that for $W/Z$-fusion. In fact, even for $\sqrt{s} = 500 \gev$ and
unpolarized beams, the Bjorken process dominates over $W$-fusion only
for relatively large Higgs masses~\cite{Djouadi:2005gi,Barger:1993wt,
Choudhury:2002qb}.

In view of this, it might seem useful to concentrate first on the
dominant channel, viz $e^+e^-\to \nu \bar\nu H$ and thereby constrain
$b_W$ and $\tilde b_W$. However, it is immediately obvious that the
Bjorken process too contributes to this final state and hence the
couplings $\Delta a_Z$, $b_Z$ and $\tilde b_Z$ have a role to play. 
Since the total rate is a $CP$-even (as well as $\tilde T$ even)
observable, it can receive contribution only from $\Re(b_V)$. 
[Note that a non-zero $\Delta a_V$ would only rescale the SM rates.] The
other non-standard couplings, odd under $CP$ and/or $\tilde T$, 
are responsible for various polar and azimuthal asymmetries and
contribute nothing to the total rate on integration over a symmetric
phase space\footnote{Remember that we work only to the linear order in 
the anomalous couplings. Thus they contribute only through the interference 
terms with the SM amplitude.}. 
This can be understood best by considering the square of the invariant
amplitude pertaining to on-shell $Z$-production, namely $e^+ e^- \to Z H$~:
\begin{equation}
\begin{array}{rcl}
|{\cal M}|^2
 & \propto & \displaystyle
|a_Z|^2 \, \frac{\ell_e^2 + r_e^2}{4} \, 
          \left[1 + \frac{E_Z^2 - p^2 \, \cos^2 \theta}{m_Z^2}\right]
\\[2ex]
& + &  \displaystyle \frac{\Re(a_Z \, b_Z^*)}{m_Z^2} \, (\ell_e^2 + r_e^2) \,
        \sqrt{s}\, E_Z
\\[2ex]
& + &  \displaystyle \frac{\Im(a_Z \, \tilde b_Z^*)}{m_Z^2} \, (\ell_e^2 - r_e^2) \,
        \sqrt{s}\, p \, \cos \theta
\end{array}
    \label{ZH_prod}
\end{equation}
with $\ell_e (r_e)$ denoting the electron's couplings to the $Z$, and
$E_Z, p, \theta$ the energy, momentum and scattering angle of the $Z$
in the c.m.-frame. The proportionality constant includes, alongwith
the couplings etc., a factor  $s / [(s - m_Z^2)^2 + \Gamma_Z^2 m_Z^2]$.
This suggests that the anomalous contribution vanishes for large $s$,
inspite of the higher-dimensional nature of the coupling.
Furthermore, Eq.(\ref{ZH_prod}) also demonstrates that neither
$\Im(b_Z)$ nor $\Re(\tilde b_Z)$ may make their presence felt if the
polarization of the $Z$ could simply be summed over. It also shows
that the contribution due to $\Im(\tilde b_Z)$ would disappear when
integrated over a symmetric part of the phase space as has been
mentioned before.  Thus, if we want to probe these couplings, we would
need to look at rates integrated only over partial (non-symmetric)
phase space. As an example, let us consider the {\em forward-backward 
asymmetry} for the $Z$-boson. As can be seen from Eq.(\ref{ZH_prod}), it
is proportional to $\Im(\tilde b_Z)$ alone. This can be understood by 
realizing that this observable is proportional to the expectation value
of $(\vec{p}_e\cdot \vec{p}_Z)$ and hence is a $CP$-odd and $\tilde
T$-even quantity just as $\Im(\tilde b_Z)$ is.  For $\Im(b_Z)$ and
$\Re(\tilde b_Z)$, which are $\tilde T$-odd, one has to look at the
azimuthal correlations of the final state fermions. Equivalently one
can look for partial cross-section, restricting the azimuthal angles
over a given range. A discussion of how partial cross-section can be used 
to probe anomalous couplings is
given in Appendix~\ref{pcs}. Further, Eq.(\ref{ZH_prod}) also
indicates that the angular distribution of the Higgs (or,
equivalently, that of the $Z$) is different for the SM piece than that
for the piece proportional to $\Re(b_Z)$; the difference getting
accentuated at higher $\sqrt{s}$. This, in principle, could be
exploited to increase the sensitivity to $\Re(b_Z)$.

In this paper we restrict ourselves to the case of a first generation linear
collider. For such $\sqrt{s}$, the interference between the $W$-fusion
diagram with the $s$-channel one is enhanced for non-zero $b_Z$ and
$\tilde b_Z$. At a first glance, it may seem that the kinematic
difference between the two set of diagrams could be exploited and
the two contributions separated from each other with some simple cuts.
However, in actuality, such a simple approach does not
suffice to adequately decouple them. It is thus contingent upon us to
first constrain the non-standard $ZZH$ couplings from processes that
involve just these and only then to attempt to use $WW$-fusion process to probe
the $WWH$ vertex.

\section{The $ZZH$ couplings}
     \label{sec:zzh}
The anomalous $ZZH$ couplings have been studied earlier in the process
$e^+e^-\to f\bar f H$ in the presence of an anomalous $\gamma\gamma H$
coupling~\cite{Hagiwara:2000tk} making use of optimal
observables~\cite{oot}.  The $CP$-violating anomalous $ZZH$ couplings
alone have also been studied in Ref.~\cite{t.han}, which constructs
asymmetries for both polarized and unpolarized beams. We, however,
choose to be conservative and restrict ourselves to unpolarized
beams. And, rather than advocating the use of complicated statistical
methods, we construct various simple observables that essentially
require only counting experiments. Furthermore, we include the decay
of the Higgs boson, account for $b$-tagging efficiencies and
kinematical cuts to obtain more realistic sensitivity limits.
%
%
Since we are primarily interested in the intermediate mass Higgs boson 
($2m_b \leq m_H \leq 140$ GeV), $H\to b\bar b$ is the dominant decay mode
with a branching fraction $\gsim 0.9$.  

\subsection{Kinematical Cuts}
For a realistic study of the process $e^+e^- \to f \bar f H (b\bar b)$,
we choose to work with a Higgs boson of mass 120 GeV and 
a collider center-of-mass energy of 500 GeV. To ensure 
detectability of the $b$-jets, we require, for each, a minimum energy and  a 
minimum angular deviation from the beam pipe. Furthermore, the two jets should 
be well separated so as to be recognizable as different ones. To be 
quantitative, we require that 
\begin{equation}
\begin{array}{rcl}
E_b, E_{\bar b} & \ge & 10 \gev, \\[1ex]
5^\circ \le \theta_b, \theta_{\bar b} & \le & 175^\circ \\[1ex]
\Delta R_{b \bar b} & \geq & 0.7
\end{array}
        \label{cuts:b}
\end{equation}
where $(\Delta R)^2 \equiv (\Delta \phi)^2 + (\Delta \eta)^2 $ with 
$\Delta \phi $ and $\Delta \eta$ denoting the separation between the two 
$b$-jets in azimuthal angle and rapidity respectively. 

For events with the $Z$ decaying into a pair of leptons or light
quarks, we have similar demands on the latter, namely
\begin{eqnarray}
E_{f}, E_{\bar f} \ge 10 \gev, \qquad
5^\circ \le \theta_{f}, \theta_{\bar f} \le 175^\circ. 
        \label{cuts:lept}
\end{eqnarray}
For leptons, i.e. $f=\ell$, we require a lepton--lepton separation:
\begin{eqnarray}
\Delta R_{\ell^- \ell^+} \geq 0.2 
\end{eqnarray}
along with a $b-$jet--lepton isolation:
\begin{eqnarray}
\Delta R_{b \ell} \geq 0.4
        \label{cuts:l_b}
\end{eqnarray}
for each of the four jet-lepton pairings. For $f=q$, i.e. light quarks, 
we impose, instead, 
\begin{equation}
\Delta R_{q_1 q_2} \geq 0.7
   \label{cuts:all_dr}
\end{equation}
for each of the six pairings.
On the other hand, if the $Z$ were to decay into neutrinos,
the requirements of Eqs.(\ref{cuts:lept}--\ref{cuts:all_dr}) are 
no longer applicable and instead we demand that the events 
contain only the two $b$-jets along with a minimum missing 
transverse momentum, viz 
\begin{eqnarray}
        p_T^{\rm miss} \geq 15 \gev \ .
   \label{cuts:miss}
\end{eqnarray}
The above set of cuts select the events corresponding to the process
of interest, rejecting most of the QED-driven backgrounds.  To further
distinguish between the role of the Bjorken diagram and that due to
the $ZZ$ ($WW$) fusion in the case of $e^+e^-H$ ($\nu \bar \nu H$)
final state we need to select/de-select the events corresponding to
the $Z$-mass pole. This is done via an additional cut on the invariant
mass of $f\bar f$, {\em viz.}
\begin{equation}
\begin{array}{rcl}
R1 &\equiv& \left| m_{f\bar f} - M_Z \right| \leq 5 \, \Gamma_Z \hspace{0.5cm}
\mbox{ select \ Z-pole} \ ,\\[1ex]
R2 &\equiv& \left| m_{f\bar f} - M_Z \right| \geq 5 \, \Gamma_Z \hspace{0.5cm}
\mbox{ de-select \ Z-pole.}
\end{array}
                \label{cuts:Z}
\end{equation}
Since an exercise such as the current one would be undertaken only after the
Higgs has been discovered and its mass measured to a reasonable accuracy, one
may alternatively demand that the energy of the Higgs (reconstructed $\bar b
b$ pair) is close to $(s + m_H^2 - m_Z^2) / (2\sqrt{s})$, namely
\begin{equation}
\begin{array}{rcl}
R1^\prime &\equiv& E^-_H \leq E_H \leq E^+_H 
\\[1ex]
R2^\prime &\equiv& E_H < E^-_H \ \mbox{or} \ E_H > E^+_H 
\end{array}
                \label{cuts:EH}
\end{equation}
where $E_H^\pm = (s + m_H^2 - (m_Z\mp5\Gamma_Z)^2) / (2\sqrt{s})$.
This has the advantage of being applicable to the
$\nu\bar\nu H$ final state as well.
The $b$-jet tagging efficiency is taken to be 0.7.
We add the statistical error and a presumed 1\% systematic error 
(accruing from luminosity measurements etc.) in quadrature. 
In other words, the fluctuation in the measurement of a cross-sections 
is assumed to be 
\begin{equation}
\Delta\sigma = \sqrt{\sigma_{SM}/{\cal L} + \epsilon^2\sigma_{SM}^2} \  \ ,
\label{Dsig}
\end{equation}
while that for an asymmetry is
\begin{equation}
(\Delta A)^2= \frac{1-A^2_{SM}}{\sigma_{SM}{\cal L}} + \frac{\epsilon^2}{2}(1-
A^2_{SM})^2. 
\label{Dasym}
\end{equation}
Here $\sigma_{SM}$ is the SM value of cross-section, ${\cal L}$ is the 
integrated luminosity of the $e^+e^-$ collider and $\epsilon$ is the 
fractional systematic error. Since we work in the linear approximation for the
anomalous couplings, any observable, rate or asymmetry, can be written as 
\[ {\cal O}(\{{\cal B}_i\}) = \sum \ O_i \ {\cal B}_i. \]
Then we define the {\em blind region} as the region in the parameter space for
which
\begin{equation}
|{\cal O}(\{{\cal B}_i\})-{\cal O}(\{0\})|\le f \ \delta{\cal O},
\end{equation}
where $f$ is the degree of statistical significance, ${\cal O}(\{0\})$ is the
SM value of ${\cal O}$ and $\delta{\cal O}$ is the statistical fluctuation in
${\cal O}$. All the limits and blind 
regions quoted in this paper are obtained using the above relation.
Note that in all the cases that we will consider, the asymmetries vanish 
identically within the SM. 
\subsection{Cross-sections}
    \label{sec:zzh_csec}
The simplest observable, of course, is the total rate. 
Note that $\Im(b_Z)$, inspite of being $\tilde T$-odd, does result in a non-zero,
though small, contribution to the total cross-section.  This is but a 
consequence of the absorptive part in the propagator and would have been 
identically zero in the limit of vanishing widths.  Since we retain
contributions to the cross-section that are at best linear in the couplings,
the major non-trivial anomalous contribution, on imposition of the $R1$ cut, 
emanates from $\Re(b_Z)$
with only subsidiary contributions from $\Im(b_Z)$.  For our
default choice (a 120 GeV Higgs at a machine operating at $\sqrt{s} =
500 \gev$), on selecting the $Z$-pole ($R1$ cut) the rates, in femtobarns,
are 
\begin{equation}
\begin{array}{rcl}
\sigma(e^+e^-) &=& 1.28 + 12.0 \; \Re(b_Z) + 0.189 \; \Im(b_Z) \\[0.5ex]
\sigma(\mu^+\mu^-) &=& 1.25 + 11.9\; \Re(b_Z)
\\[0.5ex]
\sigma(u \bar u / c \bar c) &=& 2 \ \left[4.25 + 40.2 \; \Re(b_Z)
\right]\\[0.5ex]
\sigma(d \bar d / s \bar s) &=& 2 \ \left[5.45 + 51.6 \; \Re(b_Z)
\right]
\end{array}
    \label{sig:Bj_R1}
\end{equation}
On de-selecting the $Z$-pole ($R2$ cut) we obtain, instead
\begin{eqnarray}
\sigma(e^+e^-) = \left[4.76 - 0.147 \ \Re(b_Z) \right] \fb.
\label{sig:eeR2}
\end{eqnarray}
The total rates, by themselves, may be used to put stringent constraints on
$\Re(b_Z)$. For $Z$ decaying into light quarks and $\mu$s (with the $R1$ cut)
and for an integrated luminosity of $500 \fb^{-1}$, the lack of any deviation 
from the SM expectations would give
\begin{equation}
\left|\Re(b_Z) \right| \leq 0.44 \times 10^{-2}
\label{rebz_bjork}
\end{equation}
at the $3 \sigma$ level. We do not use the $e^+e^-H$ final state 
in deriving the above constraint as it receives a contribution proportional to
$\Im(b_Z)$ too. This arises from the interference of the Bjorken 
diagram with the $ZZ$-fusion diagram due to the presence of the absorptive 
part in the near-on-shell $Z$-propagator.

\begin{figure}[!h]
\epsfig{file=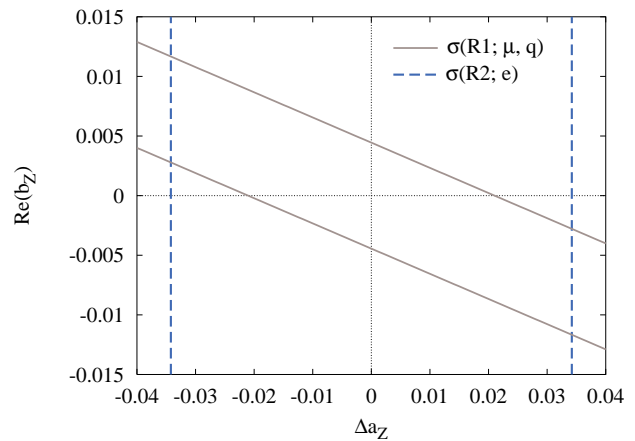,width=8.5cm}
\caption{\label{fig:az}The region in the $\Delta a_Z-\Re(b_Z)$ plane 
consistent with $3\sigma$ variations in the 
rates of Eqs.(\ref{sig:az1}) and (\ref{sig:az2}) for
an integrated luminosity of 500 fb$^{-1}$.}
\end{figure}

The cross-sections shown in Eqs.(\ref{sig:Bj_R1} \& \ref{sig:eeR2})
and the constraint of Eq.(\ref{rebz_bjork}) have been derived assuming
the SM value for $a_Z$. Clearly, any variation in $a_Z$ would affect the 
total rates and thus it is interesting to investigate possible 
correlations between $a_Z$ and $\Re(b_Z)$. Parameterizing
small variations in $a_Z$ as $a_Z = (1+\Delta a_Z)$,  
the cross-sections can be re-expressed as
\begin{eqnarray}
\sigma(R1;\mu,q) &=& \left[ 20.7 \ (1 + 2 \Delta a_Z) + 196 \ 
\Re(b_Z)\right] \fb \label{sig:az1} 
\end{eqnarray}
and
\begin{eqnarray}
\sigma(R2;e) &=& \left[4.76 \ (1 + 2 \Delta a_Z) - 0.147 \ \Re(b_Z) 
 \right] \fb \label{sig:az2}
\end{eqnarray}
where $\sigma(R1;\mu,q)$ stands for cross-section with $R1$ cut
with $\mu$ and light quarks $q$ in the final state, i.e., the combination used
to obtain Eq.(\ref{rebz_bjork}), and, as before, terms quadratic in small
parameters have been neglected. Using these two rates we can
obtain simultaneous constraints in the $\Delta a_Z-\Re(b_Z)$ plane, as
exhibited in Fig.~\ref{fig:az}. The oblique lines are obtained using
Eq.(\ref{sig:az1}) whereas the almost vertical lines are obtained
using Eq.(\ref{sig:az2}).
Thus, $\sigma(R2;e)$ alone can constrain $a_Z$ to within a few
percent of the SM value. It is amusing to consider, at this stage, the
possibility that $\Delta a_Z$ could have been large, as for example
happens in the MSSM. Clearly, the very form of Eqs.(\ref{sig:az1} \&
\ref{sig:az2}) tells us that this would have amounted to just a
two-fold ambiguity with a second and symmetric allowed region lying
around $a_Z = -1$. Since we are interested in the SM like Higgs boson, we
constrain ourselves to region near $a_Z = 1$ as shown in Fig.~\ref{fig:az}.

From Eq.(\ref{sig:Bj_R1}) and Eq.(\ref{sig:az1}) it is clear that  the total 
rates, for the $R1$ cut, depend on only one combination of $a_Z$ and
$\Re(b_Z)$, i.e. on
\begin{equation}
 \eta_1 \equiv 2 \ \Delta a_Z + 9.46 \ \Re(b_Z). 
 \label{eta1}
\end{equation}
With this combination we can write all the cross-sections given in
Eqs.(\ref{sig:Bj_R1}) and (\ref{sig:az1}) as $ \sigma = \sigma_{SM} \ ( 1 +
\eta_1)$, where the limit from Fig.~\ref{fig:az} translates to
\begin{equation}
|\eta_1| \le 0.042,
\label{eq:eta1}
\end{equation}
i.e., $\pm4.2$\% ($3\sigma$) variation of the rates. This variation can also be
parameterized with $|\Re(b_Z)| \leq 0.0044$ keeping $\Delta a_Z=0$ or
with $|\Delta a_Z|\leq 0.021$ keeping $\Re(b_Z)=0$, (i.e., the intercepts of the
solid lines in Fig.~\ref{fig:az} on the $y$-- and $x$--axes respectively).
In other words, the  individual limit, i.e. the limit
obtained keeping only one anomalous coupling non-zero, on $\Re(b_Z)$ is 0.0044
and that on $\Delta a_Z$ is 0.021. On the other hand, if the $R2$ cut were 
to be operative, we obtain the constraint $|\Delta a_Z| \leq 0.034$ almost 
independent of $\Re(b_Z)$, see Fig.~\ref{fig:az}. This constraint translates 
to a $\pm6.8$\% ($3\sigma$) variation in the rate $\sigma(R2;e)$. 

Note that the contributions proportional to the absorptive part of the 
$Z$-propagator are proportional to $\Gamma_Z$ away from the resonance, i.e. 
for the $R2$ cut. Hence in this case it is a higher order effect and thus 
ignored in Eqs.(\ref{sig:eeR2}) and (\ref{sig:az2}).
On the other hand, near the $Z$-resonance these terms are proportional to
$1/\Gamma_Z$ and hence of the same order in the perturbation theory. 
Thus, in order to be consistent at a given
order in the coupling $\alpha_{em}$ we retain these contribution only with 
the $R1$ cut.
\subsection{Forward-backward asymmetry}
    \label{sec:zzh_FB}
The final state constitutes of two pairs of identifiable particles :
$b\bar b$ coming from the decay of Higgs boson and $f\bar f$, where
$f\neq b$. One can define forward-backward asymmetry with respect to
all the four fermions. But we choose, among them, the asymmetries with
definite transformation properties under $CP$ and $\tilde T$. In the
present case we have only one such forward-backward asymmetry,
i.e. the expectation value of $(\vec{p}_{e^-} - \vec{p}_{e^+})\cdot
(\vec{p}_{f} + \vec{p}_{\bar f})$. In other words, it is the
forward-backward asymmetry with respect to the polar angle of the
Higgs boson (up to an overall sign) and given as
\begin{equation}
A_{FB}(\cos\theta_H) = \frac{\sigma(\cos\theta_H>0) - \sigma(\cos\theta_H<0)}
{\sigma(\cos\theta_H>0) + \sigma(\cos\theta_H<0)}.
\label{A_FB}
\end{equation}
This observable is $CP$ odd and $\tilde T$ even, and hence a 
probe purely of 
of $\Im(\tilde b_Z)$ [see Table \ref{table:prop}]. Note that this
asymmetry is proportional to $(r_e^2-l_e^2)$, where $r_e \, (l_e)$
are the right-(left-) handed couplings of the electron to the $Z$-boson.
With the $R1$ cut---see Eq.(\ref{cuts:Z})---operative, a
semi-analytical expression for this asymmetry, keeping only terms linear  in
the anomalous couplings, is given by
\begin{equation}
A_{FB}(c_H) = 
\left\{
\begin{array}{lcl} 
\displaystyle
\frac{ 0.059 \ \Re(\tilde b_Z) -1.22 \ \Im(\tilde b_Z)}{1.28 }
&  & (e^+e^-)\\[1.5ex]
\displaystyle
\frac{-1.2 \ \Im(\tilde b_Z)}{1.25 }
&  & (\mu^+\mu^-) \\[1.5ex]
\displaystyle
\frac{-18.5 \ \Im(\tilde b_Z)}{19.4} &  & 
(q \bar q)
\end{array}
\right.
\end{equation}
In the above, ``$q$'' stands for all four flavors of light quarks
summed over and $c_H \equiv \cos\theta_H$.  Note here, that the
contribution to the denominator of Eq.(\ref{A_FB}) from the anomalous
terms have been dropped as the formalism allows us to retain terms
only upto the first order in these couplings. In any case, their
presence would have had only a miniscule effect on the ensuing bounds.
The asymmetry corresponding to the $R2$ cut is very small and is not
considered.  Omitting the ($e^+ e^-$) final state on account
of the presence of $\Re(\tilde b_Z)$, we use only the light quarks and
$\mu$s.  For an integrated luminosity of 500 fb$^{-1}$, the
corresponding $3\sigma$ limit is
\begin{equation}
|\Im(\tilde b_Z)|\leq0.038.
\label{lim:imbtz}
\end{equation}
\begin{figure}[!h]
\epsfig{file=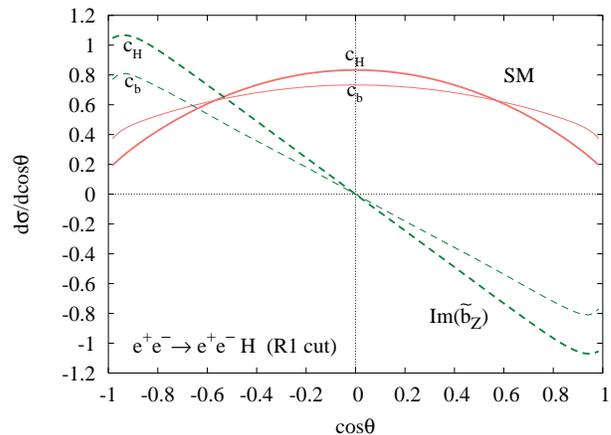,width=8.5cm}
\caption{\label{fig:cH}Polar angle distribution of the $b-$quark (thin line)
follows that of Higgs boson (thick line) but is a bit smeared out. 
This is because the 
$b-$quark is spherically distributed in the rest frame of Higgs boson.}
\end{figure}

It is interesting to speculate as to the sensitivity of the a
forward-backward asymmetry constructed with respect to the angle
subtended by, say, the $b$-jet, rather than that for the reconstructed
$H$.  Since the $b-$quarks are spherically distributed in the rest
frame of Higgs, their angular distributions track that of the Higgs
boson, modulo some smearing [see Fig.~\ref{fig:cH}].  This is as true
for the anomalous contribution as for the SM.  The smearing only
serves to decrease the sensitivity as is evinced by the
forward-backward asymmetries constructed with respect to $\theta_b$, 
the polar angle of the $b-$quark. For events corresponding to the $R1$
cut, this amounts to
\[
A_{FB}(c_b) = 
\left\{
\begin{array}{lcl} 
\displaystyle
\frac{0.0489 \ \Re(\tilde b_Z) - 0.909 \ \Im(\tilde b_Z)}
 {1.28 }
&  & (e^+e^-)\\[1.5ex]
\displaystyle
\frac{-0.892 \ \Im(\tilde b_Z)}{1.25 }
&  & (\mu^+\mu^-) \\[1.5ex]
\displaystyle
\frac{-13.8 \ \Im(\tilde b_Z)}{19.4}&  & 
(q \bar q)
\end{array}
\right.
\]
Using final states with $\mu$s or light quarks,  one may then use 
$A_{FB}(c_b)$ to probe down to
\[
|\Im(\tilde b_Z)| \leq 0.051   
\]
at $3\sigma$ level, for an integrated luminosity of 500 fb$^{-1}$, and
assuming all the other anomalous couplings to be zero. It is not surprising 
that the sensitivity is lower as compared to that of $A_{FB}(c_H)$---see 
Eq.\ref{lim:imbtz}---for $\theta_b$ carries only subsidiary 
information leading to a reduction in the size of the asymmetry. Put 
differently, the distribution for the $b$ is identical to 
that for the $\bar b$ (thereby eliminating the need 
for charge measurement) and each is driven primarily by $\theta_H$.

Note that we have desisted from using the non-zero forward-backward asymmetry
in the polar angle distribution  of $f(\bar f)$. 
Such observables do not have the requisite $CP$-properties and 
are, in fact, non-zero  even within the SM. The
presence of non-zero $\Im(\tilde b_Z)$ provides only an additional source for
the same and the limits extractable would be weaker 
than those we have obtained.
\subsection{Up-down asymmetry}
The Higgs being a spin--0 object, its decay products are isotropically
distributed in its rest frame. This, however, is not true of the $Z$. Still,
$CP$ conservation ensures that the leptons from $Z$ decay are symmetrically
distributed about the plane of production. Thus, an up-down asymmetry defined as
\begin{equation}
A_{UD}(\phi) = \frac{\sigma(\sin\phi>0) - \sigma(\sin\phi<0)}
{\sigma(\sin\phi>0) + \sigma(\sin\phi<0)}
\label{A_UD}
\end{equation}
can be non-zero for the anomalous couplings.
In other words, a non-zero expectation value for $\left[ (\vec{p}_{e^-} -
  \vec{p}_{e^+}) \times \vec{p}_H \right] \cdot
(\vec{p}_{f}-\vec{p}_{\bar f})$ is a $CP$ odd (and $\tilde{T}$ odd)
observable. In our notation, it is driven by the non-zero real part of
$\tilde b_Z$, and, for our choice of parameters, amounts to
\begin{equation}
\begin{array}{rcl}
A_{UD}(\phi_{e^-})&=& \displaystyle 
\frac{-0.354 \ \Re(\tilde b_Z) - 0.226 \ \Im(\tilde b_Z)}{1.28}
\\[2ex]
A_{UD}(\phi_{\mu^-})&=& \displaystyle 
\frac{-0.430  \ \Re(\tilde b_Z)}{1.25  } \\[2ex]
A_{UD}(\phi_{u})&=&\displaystyle \frac{-4.62  \ \Re(\tilde b_Z)}{4.25 } \\[2ex]
A_{UD}(\phi_{d})&=&\displaystyle \frac{-7.98  \ \Re(\tilde b_Z)}{5.45 }
\end{array}
     \label{eq:aud_numbers}
\end{equation}
up to linear order in the anomalous couplings. 
In obtaining Eq.\ref{eq:aud_numbers}, the $R1$ cut has been imposed 
on the $f\bar f$ invariant mass. 
Note that, except for the $e^+e^-H$ final state, $A_{UD}$ is a 
probe purely of $\Re(\tilde b_Z)$. The cross section for
the $e^+e^-H$ final state receives additional contribution from
$\Im(\tilde b_Z)$ due to the absorptive part of the $Z$-propagator in
the Bjorken diagram. 
Although $A_{UD}(\phi_u)$ and $A_{UD}(\phi_d)$ offer much larger 
sensitivity to $\Re(\tilde b_Z)$ than do either of $A_{UD}(\phi_{e^-})$
and $A_{UD}(\phi_{\mu^-})$, the former can not be used as the 
measurement of such asymmetries requires charge determination 
for light quark jets. However, one can
determine the charge of $b$-quarks~\cite{b-charge} and using
$A_{UD}(\phi_b)$, for the $b$'s resulting from the $Z$ decay, 
we may obtain, for an integrated luminosity of 500 fb$^{-1}$,
a $3\sigma$ bound of 
\[ |\Re(\tilde b_Z)|\leq0.042 \]
with 100\% charge determination efficiency and only 
\[ |\Re(\tilde b_Z)|\leq0.089 \]
if the efficiency were 20\%. Note though that the $Z\to b\bar b$ final
state is beset with additional experimental complications (such as
final state combinatorics) than the semileptonic channels and hence we
would not consider this in deriving our final limits.  
Note that we do not account for any combinatorics in obtaining the above said
limits. However, we argue that the invariant masses of the $b\bar b$ pair
coming from $Z$ and $H$ are non-overlapping hence the change in the above limits
due to combinatorics are expected to be small.

Another obvious observable is $A_{UD}(\phi_{\mu^-})$. Using this, an 
integrated luminosity of 500 fb$^{-1}$, would lead to a $3\sigma$ constraint of
\begin{equation}
\Re(\tilde b_Z)|\leq0.35 .
\label{r1:rebtz}
\end{equation}
The reduced sensitivity of the $Z\to \mu^+ \mu^-$ channel
as compared to the $Z \to b \bar b$ channel is easy to understand. As 
Eq.(\ref{mesq_quark}) demonstrates, 
$A_{UD}(\phi_{f}) \propto  (r_e^2 - \ell_e^2) \, (r_f^2 - \ell_f^2)$. 
Since $|r_\mu| \approx |\ell_\mu|$, this
naturally leads to an additional suppression for $A_{UD}(\phi_{\mu^-})$. 

For the $e^+ e^- H$ case, on the other hand, the $ZZ$-fusion diagram 
leads to a contribution that is proportional to $(r_e^2 + \ell_e^2)^2$ 
and is, thus, unsuppressed. Accentuating this contribution by employing 
the $R2$ cut on $m_{e e}$, we have 
\begin{eqnarray}
A_{UD}^{R2}(\phi_{e^-})
&=&\frac{5.48 \ \Re(\tilde b_Z) } {4.76}
\end{eqnarray}
and this, for an integrated luminosity of 500 fb$^{-1}$, leads to a
$3\sigma$ constraint of
\begin{eqnarray}
|\Re(\tilde b_Z)|&\leq&0.057.
\end{eqnarray}
Here we note that the limit on $\Re(\tilde b_Z)$, obtained using the 
$R2$ cut given above, is 
much better than the one obtained using $R1$ cut in Eq.(\ref{r1:rebtz}), 
or even the one derived from the $4 \, b$ final state assuming a 
$20\%$ charge detection efficiency.
\subsection{Combined polar and azimuthal asymmetries}
Rather than considering individual asymmetries involving the (partially
integrated) distributions in either of the polar or the azimuthal
angle, one may attempt to combine the information in order to
potentially enhance the sensitivity. To this end, we define a momentum
correlation of the form 
\begin{eqnarray}
{\cal C}_1 &=& 
\left[(\vec{p}_{e^-} - \vec{p}_{e^+})\cdot\vec{p}_{\mu^-}\right] 
\nonumber\\
&&\left[\left[ (\vec{p}_{e^-} - \vec{p}_{e^+}) \times \vec{p}_H
 \right]\cdot (\vec{p}_{\mu^-}-\vec{p}_{\mu^+})\right],
\end{eqnarray}
where the sign of the term in the first square bracket decides if the $\mu^-$
is in forward($F$) hemisphere with respect to the
direction of $e^-$ or backward($B$).
Similarly, the sign of the term in second square bracket defines if $\mu^-$ is
above($U$) or below($D$) the Higgs production plane. Thus the expectation value
of the sign of this correlation is same as the combined polar-azimuthal
asymmetry given by,
\begin{eqnarray}
A(\theta_\mu,\phi_\mu)&=&\frac{(FU)+(BD)-(FD)-(BU)}{(FU)+(BD)+(FD)+(BU)} 
\nonumber\\
&=&\frac{ 0.659 \; \Im(b_Z) - 0.762 \; \Re(\tilde{b}_Z)}{1.25}
\label{amix}
\end{eqnarray}
with the second equality being applicable for the $R1$ cut. In the
above, $(FU)$ is the partial cross-sections for $\mu^-$ in the
forward-up direction and so on for others. Note that ${\cal C}_1$ is
$\tilde T$-odd but does not have a definite $CP$ and hence depends on
both the $\tilde T$-odd couplings as seen in Eq.(\ref{amix}).  We do
not consider the analogous asymmetry for $q\bar q H$ final state as it
demands charge determination for light quarks (although $Z \to b
\bar b$ may be considered profitably).  Similarly, for the $e^+e^-H$
final state, $A(\theta_e,\phi_e)$ receives contributions from $\tilde
T$-even couplings as well and hence not considered for the analysis.

\begin{figure}[!h]
\epsfig{file=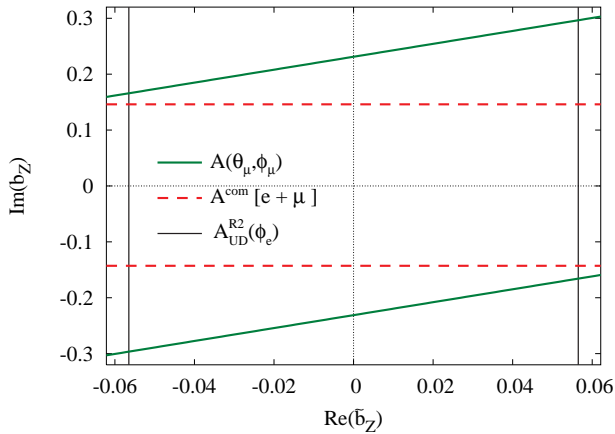,width=8.5cm}
\caption{\label{fig:t-odd}Region in $\Re(\tilde b_Z)-\Im(b_Z)$ plane
corresponding to the $3\sigma$ variation of asymmetries. Slant lines are for
 $A(\theta_\mu,\phi_\mu)$ and the vertical lines are for $A_{UD}$ with an
 integrated luminosity of 500 fb$^{-1}$.
The horizontal limit shown are
due to $A^{com}$ for $e^-$ and $\mu^-$ in the final state.}
\end{figure}
Using this $\tilde T$-odd asymmetry along with $A_{UD}$,
$\Re(\tilde b_Z)$ and $\Im(b_Z)$ can be constrained simultaneously.
Fig.~\ref{fig:t-odd} shows the limit on $\Im(b_Z)$ as a function of
$\Re(\tilde b_Z)$. 
\subsection{Another asymmetry}
Similar to the previous subsection, we can define a $CP$-even and 
$\tilde T$-odd correlation as
\begin{eqnarray}
{\cal C}_2 &=& \left[(\vec{p}_{e^-} - \vec{p}_{e^+})\cdot\vec{p}_{Z}\right] 
\nonumber\\
&&\left[\left[ (\vec{p}_{e^-} - \vec{p}_{e^+}) \times \vec{p}_H
 \right]\cdot (\vec{p}_{f}-\vec{p}_{\bar f})\right]
\end{eqnarray}
which is a probe of the $CP$-even and $\tilde T$-odd coupling
$\Im(b_Z)$. Here, the sign of the term in the first square bracket
decides whether the Higgs boson is in the forward($F'$) or
backward($B'$) hemisphere, while the sign of the term in the second
square bracket indicates if $f$ is above($U$) or below($D$) the Higgs
production plane. The expectation value of the sign of ${\cal C}_2$
can thus be expressed as an  asymmetry of the form
\begin{eqnarray}
A^{com} = \frac{(F'U)+(B'D)-(F'D)-(B'U)}{(F'U)+(B'D)+(F'D)+(B'U)}.
\label{Acom}
\end{eqnarray}
The semi-analytical expression for this asymmetry for $R1$ cut is given by
\begin{eqnarray}
A^{com}_\mu&=&\frac{ 0.766 \; \Im(b_Z)}{1.25 }\nonumber \\[1.5ex]
A^{com}_e&=&\frac{ 0.757 \; \Im(b_Z) - 0.048 \; \Re(b_Z)}{1.28} \\[1.5ex]
A^{com}_b&=&\frac{ 14.2  \; \Im(b_Z)}{5.45} \nonumber
\end{eqnarray}
Using final states with either electrons or muons, we would obtain 
a $3\sigma$ limit of
\begin{equation}
|\Im(b_Z)| \le 0.14,
\end{equation}
for 500 fb$^{-1}$ of integrated luminosity and maintaining 
all the other form-factors to be zero. 
This limit is better than the one obtained in the preceding subsection. 

Once again, inclusion of the $Z \to b \bar b$ channel, i.e.  a
measurement of $A^{com}_b$, would improve the situation dramatically
even for a nominal charge detection efficiency. For example, an
efficiency as low as just 20\%, is enough to obtain $$|\Im(b_Z)| \le
0.050.$$ We reemphasize though that our final results do not
exploit this possibility.

We also note that the sensitivity to the $\tilde T$-odd couplings is
large for $f=q$ as compare to $f=\ell$. As the expressions in the
Appendix~\ref{M2} demonstrate, $A_{UD}$, with the $R1$ cut operational, is
proportional to $(l_e^2-r_e^2)(l_f^2-r_f^2)$ and $A^{com}$ is
proportional to $(l_e^2+r_e^2)(l_f^2-r_f^2)$. Thus, for $f=\ell$, the
asymmetries are proportional to at least one power of $(l_e^2-r_e^2)$
where $|r_e|\approx |l_e|$ and hence are smaller compared to those for the 
$f=q$ case.

\subsection{Summary of Limits on the $ZZH$ couplings}
In the preceding five subsections we discussed observables which will
be able to probe each of the five anomalous $ZZH$ couplings. The
ensuing limits are summarized in Table~\ref{tab:ind-lim}.
\begin{table}
\caption{\label{tab:ind-lim}Limits on anomalous $ZZH$ couplings from
various observables at $3\sigma$ level at an integrated luminosity of 500 
fb$^{-1}$.}
\begin{ruledtabular}
\begin{tabular}{cll}
Coupling & $3 \sigma$ Bound & Observable used\\ \hline
$|\Delta a_Z|$ & $0.034$ & $\sigma$ with $R2$ cut; $f=e^-$\\
$|\Re(b_Z)|$  &
$\left\{ \begin{array}{l}0.0044 \cr 
                        \hspace*{0ex} (\Delta a_Z = 0) \cr
                        0.012 \cr
                        \hspace*{0ex} (|\Delta a_Z| = 0.034) \cr
         \end{array} \right.$  
&$\sigma$ with $R1$ cut; $f=\mu,q$\\
$|\Im(b_Z)|$ &$0.14$ & $A^{com}$ with $R1$ cut; $f=\mu^-,e^-$\\
$|\Re(\tilde b_Z)|$ & $0.057$ & $A_{UD}(\phi_{e^-})$ with $R2$ cut\\
$|\Im(\tilde b_Z)|$ & $0.038$ & $A_{FB}(c_H)$ with $R1$ cut; \\
& & $f=\mu,q$
\end{tabular}
\end{ruledtabular}
\end{table}

Several points are in order here
\begin{itemize}
\item Recall that, of the five anomalous terms, only two 
viz $\Delta a_Z$ and $\Re(b_Z)$,  have identical transformations under 
both $CP$ and $\tilde T$. Consequently, the contributions proportional 
to the two are  intertwined and can only be partially separated. 
In fact, the most general limits on these two are to be obtained from
Fig.~\ref{fig:az}. 

\item As for the other three couplings, we have been able to construct
observables that are sensitive to only a single coupling, thereby
disengaging each of the corresponding bounds in
Table~\ref{tab:ind-lim} from contaminations from any of the other
couplings.

\item The polar-azimuthal asymmetry, $A(\theta,\phi)$, is sensitive to
$\tilde T$-odd couplings. However, the limits obtained using $\mu$
final state alone are weaker than the ones obtained by combining
$A_{UD}^{R2}$ and $A^{com}$ (see Fig~\ref{fig:t-odd}). Inclusion of
electrons in the final state will improve the sensitivity of
$A(\theta,\phi)$,  but only at the cost of contamination by the $\tilde
T$-even $ZZH$ couplings. Thus, in our present analysis, the role of
$A(\theta,\phi)$ is only a confirmatory one.

%

\item Note,  however, that many of these asymmetries are proportional 
to $(l_e^2-r_e^2) = (1-4 \, \sin^2\theta_W)$. Since this parameter 
is known to receive large radiative corrections, the importance of 
calculating higher-order effects cannot but be under-emphasized.

\item Observables constraining $\tilde T$-odd
couplings required charge determination of fermion $f$ in the $f\bar f H(b\bar
b)$ final state thereby eliminating (the dominant) 
$f=q$ final states from the analysis.
This explains a relatively poor limit on $\Im(b_Z)$. For $f=\nu$, the process
involves $WWH$ couplings as well. This is discussed in the next section.
\end{itemize}

\section{The $WWH$ couplings}
\label{sec:wwh}
As discussed at the beginning of the last section, the contribution
from non-standard $ZZH$ couplings to the $\nu\bar\nu H(b\bar b)$ final
state is not negligible even if on-shell $Z$ production is disallowed
by imposing the aforementioned $R2$ cut. With the neutrinos being
invisible, we are left with only two observables: the total
cross-section and the forward-backward asymmetry with respect to the
polar angle of the Higgs boson. The deviation from the SM expectations
for the cross section depends mainly on $\Delta a_V$ and
$\Re(b_V)$. Similarly, the forward-backward asymmetry can be
parameterized, in the large, by just $\Im(\tilde{b}_V)$. The
contribution of the other couplings, viz. $\Im(b_V)$ and
$\Re(\tilde{b}_V)$, to either of these observables are proportional to
the absorptive part of $Z$-propagator and are understandably
suppressed, especially for the $R2$ cut.

Now, irrespective of the $CP$ properties of the Higgs, its decay
products are always symmetrically distributed in its rest
frame. In addition, the momentum of the individual neutrino is not
available for the construction of any $\tilde T-$odd
asymmetry. Consequently, we do not have a direct probe of $\Im(b_V)$ and
$\Re(\tilde{b_V})$, i.e. the $\tilde T-$odd couplings.

\begin{figure}[!ht]
\epsfig{file=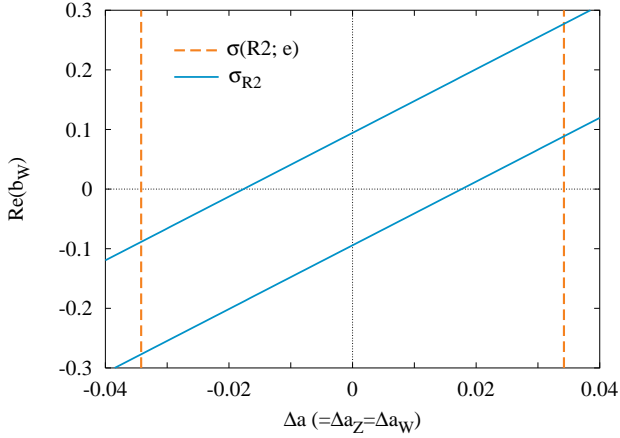,width=8.5cm}
\caption{\label{fig:rebw-k2}Region of $\Delta a - \Re(b_W)$ plane
corresponding to $3\sigma$ variation of $\sigma_{2}$ for $L=500$
fb$^{-1}$.  The vertical line shows the limit on $\Delta a_Z$ from
Fig.~\ref{fig:az}.}
\end{figure}

The event selection criteria we use are the same as in previous
section except that the cuts of Eqs.~(\ref{cuts:lept} \&
\ref{cuts:l_b}) are replaced by that of Eq.~(\ref{cuts:miss}).
Imposing $\Delta a_W =~\Delta a_Z~\equiv~\Delta a$ as argued for
earlier, the resultant cross-section, for the $R1^\prime$ cut, can be
parameterized as,
\begin{eqnarray}
\sigma_{1} &=& \left[7.69 \ (1+ \eta_1) - 1.89 \ \Im(b_Z) 
\right.\nonumber\\
&& \left. + 0.458 \ \Re(b_W) + 0.786 \ \Im(b_W) \right] \fb 
\label{sig:nn1}
\end{eqnarray}
while the $R2^\prime$ cut would lead to
\begin{eqnarray}
\sigma_{2} &=& \left[52.1 \ (1+ 2 \ \Delta a) - 6.99 \ \Re(b_Z) 
 \right.\nonumber\\
&& \left. - 0.162 \, \Im(b_Z) - 19.5 \ \Re(b_W) \right] \fb 
\label{sig:nn2}
\end{eqnarray}
Note here that the same $\eta_1=2 \ \Delta a + 9.46 \ \Re(b_Z)$ as defined in
Eq.(\ref{eta1}) appears above owing to the assumption $\Delta a_W =~\Delta
a_Z~\equiv~\Delta a$.
As the contributions proportional to $\Im(b_V)$ appear due to
interference of the $WW-$fusion diagram with the absorptive part of
the $Z$-propagator in Bjorken diagram, formally, these terms are at
one order of perturbation higher than the rest. Note that the bounds
in Table~\ref{tab:ind-lim} imply that
\[
| 6.99 \ \Re(b_Z) + 0.162 \, \Im(b_Z) | \leq 0.0839
\]
and hence the corresponding contribution to $\sigma_{2}$ is at the
per-mille level. Since we are not sensitive to such small
contributions, we may safely ignore this combination for all further
analysis.  Looking at fluctuations in $\sigma_{2}$, the $3\sigma$
bound would, then,  be
\begin{equation}
| 2 \ \Delta a - (19.5/52.1) \Re(b_W)  | \leq 0.035.
\label{rebw_lim}
\end{equation}
The limits on $\Re(b_W)$ and $\Delta a$ are thus strongly correlated
and displayed in Fig.\ref{fig:rebw-k2}. Note that a complementary bound 
on $\Delta a$ had already been obtained in 
Section \ref{sec:zzh_csec}. If we assume that, of these two couplings, only 
one is non-zero, the corresponding individual limits would be 
$|\Re(b_W)| \leq 0.097$ (if $\Delta a=0$) and, similarly,
 $|\Delta a| \leq 0.017$ (if $\Re(b_W)=0$). Interestingly, the last mentioned 
bound is twice as strong as that obtained in Section \ref{sec:zzh_csec}. 
Of course, had we not made the assumption of $\Delta a_W = \Delta a_Z$, 
or made a different assumption, the bounds derived above 
(and Fig.\ref{fig:rebw-k2}) would have looked very different.

\begin{figure}[!ht]
\epsfig{file=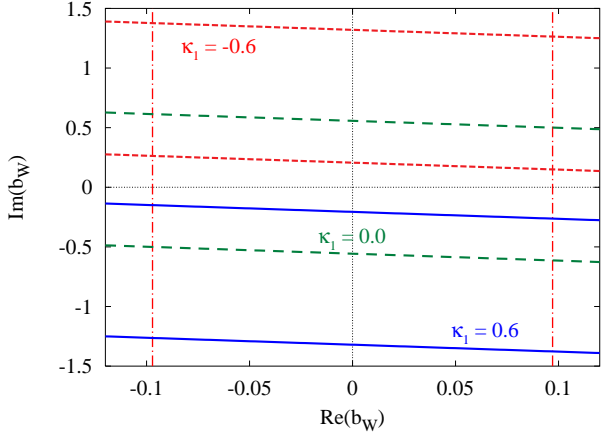,width=8.5cm}
\caption{\label{fig:bw}Region of $\Im(b_W)-\Re(b_W)$ plane corresponding to
$3\sigma$ variation of $\sigma_{1}$ for $\kappa_1$=0.0 (big-dashed line), 
0.6 (solid line) and $-0.6$ (small-dashed line). Vertical lines show limit on 
$\Re(b_W)$ obtained from Fig.~\ref{fig:rebw-k2} for $\Delta a=0$.}
\end{figure}

Having constrained $\Re(b_W)$, we may now use  $\sigma_{1}$ 
to investigate possible bounds on $\Im(b_W)$. To this end, it is useful to 
define a further subsidiary variable $\kappa_1$ as 
\begin{equation}
\begin{array}{rcl}
\kappa_1 &\equiv& \displaystyle 7.69 \ \eta_1 - 1.89 \ \Im(b_Z), \\[1ex]
|\kappa_1| & \leq & 0.604
\end{array}
\end{equation}
with the inequality having been derived using
Table~\ref{tab:ind-lim}. The corresponding constraint in the
$\Re(b_W)$--$\Im(b_W)$ plane is shown in Fig.~\ref{fig:bw} for various
representative values of $\kappa_1$. Clearly, the contamination from
the $ZZH$ couplings is very large and inescapable. Any precise
measurement of $\Im(b_W)$, in the present case, requires very accurate
determination of the $ZZH$ vertex.  The individual limit on $\Im(b_W)$,
for $\kappa_1=\Re(b_W)=0$ is given in Table~\ref{tab:lim-wwh}.

\begin{figure}
\epsfig{file=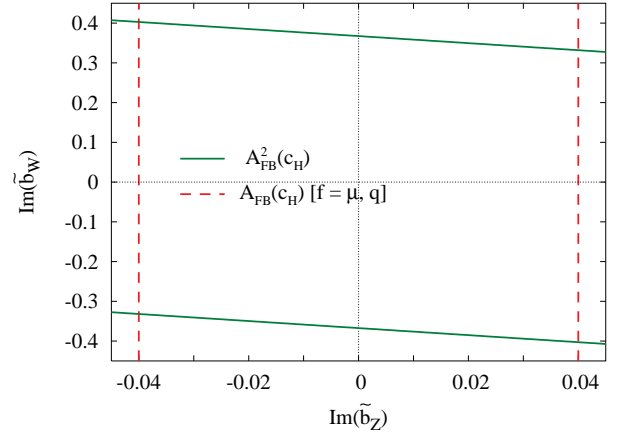,width=8.5cm}
\caption{\label{fig:A2}Regions of $\Im(\tilde b_Z)-\Im(\tilde b_W)$ plane
corresponding to $3\sigma$ variation of $A^2_{FB}(c_H)$. The vertical line 
denote limits on $\Im(\tilde b_Z)$ from Table~\ref{tab:ind-lim}.}
\end{figure}

Next, we look at the forward-backward asymmetry with respect to $c_H$
which, for our cuts, we find to be
\begin{eqnarray}
A_{FB}^1(c_H) &=&\left[-1.20 \ \Re(\tilde b_Z) - 7.11 \ \Im(\tilde b_Z) 
 \right.\nonumber\\
&&\left.+ 0.294 \ \Re(\tilde b_W) - 0.242 \ \Im(\tilde b_W)\right]/7.69 \hspace{0.5cm}
\label{asm:nnh1}
\end{eqnarray}
for the $R1^\prime$ cut, while for the $R2^\prime$ cut it is
\begin{eqnarray}
A_{FB}^2(c_H) &=&\left[ 3.55 \ \Im(\tilde b_Z) + 4.00 \ \Im(\tilde b_W)
\right]/52.1
\label{asm:nnh2}
\end{eqnarray}
Clearly $A_{FB}^2$ is the one that is more sensitive to $\Im(\tilde
b_W)$.  Once again, there is a strong correlation with $\Im(\tilde
b_Z)$, which of course has already been constrained (in Section
\ref{sec:zzh_FB}) from a consideration of similar forward-backward
asymmetries in the $b \bar b q \bar q$ and $b \bar b \mu^- \mu^+$
channels.  The resultant constraint in the $\Im(\tilde
b_W)$--$\Im(\tilde b_Z)$ plane is displayed in Fig.~\ref{fig:A2}.  And
assuming a vanishing $\Im(\tilde b_Z)$, the individual limit on
$\Im(\tilde b_W)$ is 0.37.
\begin{table}[tb]
\caption{\label{tab:lim-wwh}Individual limits on anomalous $WWH$ couplings from
various observables at $3\sigma$ level at an integrated luminosity of 500 
fb$^{-1}$}
\begin{ruledtabular}
\begin{tabular}{ccll}
Coupling & &Limit & Observable used\\ \hline
$|\Delta a|$ &$\leq$ & $0.017$& $\sigma_{2}$ \\
$|\Re(b_W)|$ &$\leq$& $0.094$ & $\sigma_{2}$ \\
$|\Im(b_W)|$ &$\leq$& $0.56$ & $\sigma_{1}$ \\
$|\Re(\tilde b_W)|$ &$\leq$& $1.4 $ & $A_{FB}^1(c_H)$\\
$|\Im(\tilde b_W)|$ &$\leq$& $0.37$ & $A_{FB}^2(c_H)$\\
\end{tabular}
\end{ruledtabular}
\end{table}

\begin{figure}[!ht]
\epsfig{file=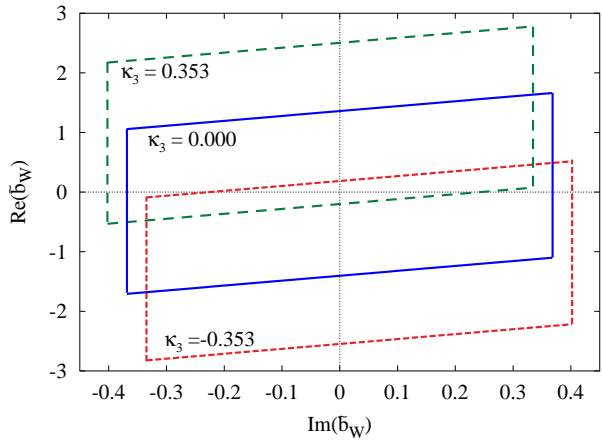,width=8.5cm}
\caption{\label{fig:A1}Regions of $\Re(\tilde b_W)-\Im(\tilde b_W)$ plane
corresponding to $3\sigma$ variation of $A^1_{FB}(c_H)$
for $\kappa_3 = 0$(solid lines), 0.353 (big-dashed lines) and -0.353
(small-dashed lines). The vertical line is the limit on $\Im(\tilde b_W)$ from 
Fig~\ref{fig:A2}.}
\end{figure}

The only coupling that remains to be constrained at this stage 
is $\Re(\tilde b_W)$. While $A^1_{FB}$ does depend on this parameter, 
it, unfortunately, also depends on the other three 
$CP$-odd anomalous coupling as well. However, since the lack of sufficient 
kinematic variables prevent us from constructing another $CP$-odd 
observables, we are forced to use $A^1_{FB}$ alone, despite its 
low sensitivity to $\Re(\tilde b_W)$. 
Collecting all the relevant $ZZH$ vertex dependence into one variable 
by defining
\begin{equation}
\kappa_3 \equiv 1.20 \ \Re(\tilde b_Z) + 7.11 \ \Im(\tilde b_Z) \ ,
\end{equation} we have
\[
\begin{array}{rcl}
A_{FB}^1(c_H) &=& \displaystyle
\left[ - \kappa_3 + 0.294 \ \Re(\tilde b_W) - 0.242 \ \Im(\tilde b_W) \right]
/ 7.69 \\[2ex]
|\kappa_3| & \leq & 0.353 \ .
\end{array}
\]
The inequality is a consequence of the bounds derived in the 
previous section (see Table.\ref{tab:ind-lim}). 
In Fig.~\ref{fig:A1}, we show the limit on $\Re(\tilde b_W)$ as 
a function of $\Im(\tilde b_W)$ for three representative values 
of $\kappa_3$, namely $\kappa_3 = 0.00,\pm0.353$. The most general limit on 
$\Re(\tilde b_W)$ can then be gleaned from this figure.

\begin{table}[!ht]
\caption{\label{tab:com-wwh}Simultaneous limits on anomalous $WWH$ couplings 
from various observables at $3\sigma$ level at an integrated luminosity of 
500 fb$^{-1}$.}
\begin{ruledtabular}
\begin{tabular}{ccll}
Coupling & &$\Delta a=0$& $\Delta a\neq0$\\ \hline
$|\Delta a|$ &$\leq$& -- & 0.034 \\
$|\Re(b_W)|$ &$\leq$& 0.097 & 0.28 \\
$|\Im(b_W)|$ &$\leq$& 1.4  & 1.4  \\
$|\Re(\tilde b_W)|$ &$\leq$& 2.8 & 2.8 \\
$|\Im(\tilde b_W)|$ &$\leq$& 0.40 & 0.40
\end{tabular}
\end{ruledtabular}
\end{table}
Note that, unlike in the case of the $ZZH$ couplings, we have largely
been unable to construct observables that are primarily dependent only
on a given anomalous coupling. In other words, the constraints are
correlated.  Thus, it is of interest to obtain the maximal size that
these couplings may assume with the aid of such correlations. Such an 
analysis may be performed by examining the 9-dimensional parameter space
(i.e., 
both $ZZH$ and $WWH$ couplings) and delineating the part 
that would be consistent with {\em all} the observables to a 
given level of confidence. Clearly, the lack of correlations for the 
$ZZH$ couplings renders the {\em blind region} to be trivial 
in five of the nine dimensions and the extent of these remain the 
same as in Table~\ref{tab:ind-lim}. The most general 
simultaneous limits on the anomalous $WWH$ couplings obtained using
this method are presented in Table~\ref{tab:com-wwh} for both $\Delta a=0$
and $\Delta a\neq0$. For the $\tilde T$-even couplings, 
such limits are comparable to the corresponding individual limits
(Table~\ref{tab:lim-wwh}) with only a small dilution due to
contamination from $ZZH$ vertex. For the $\tilde T$-odd couplings,
however, the lack of any $\tilde T$-odd observable results 
in a possibly large contamination from
$ZZH$ vertex.  Consequently, the limits on $\Im(b_W)$ and
$\Re(\tilde b_W)$ are only ``indirect'' and hence poor.
Finally, the effect of a non-zero $\Delta a$ is seen only in 
$\Re(b_W)$ due to the large correlation between them 
(see Fig.~\ref{fig:rebw-k2}).

\section{Discussions}
\label{sec:res}
We have constructed observables which, due to their $CP$ and $\tilde
T$ transformation properties, receive contributions only from specific
anomalous couplings with matching $CP$ and $\tilde T$ properties. Thus, 
most of the observables we construct are sensitive only to a single
anomalous coupling.  This one-to-one correspondence between the
observable and the anomalous coupling allows us to obtain a robust
constraint on the latter, {\it independent} of the values of all the
other anomalous couplings.  Thus we see from Tables~\ref{tab:ind-lim}
-- \ref{tab:com-wwh} that the individual and simultaneous limits are
the same (or very similar) in most cases. The observables we construct are
also very simple from the point of view of experimental measurements.
In other words,  they are both very physical and easily
implementable in actual experiments.

It should also be noted that the limits that we quote on
the anomalous couplings, other than those on $\Re (b_V)$, are obtained
using only asymmetries. In general, asymmetries are more robust
with respect to the effects of radiative corrections, except in
situations where the tree level contributions are accidentally
small. This also means that we have a clear indication as to which
observables bear a tighter scrutiny while assessing the effect of
radiative corrections. Available calculations of higher order 
corrections, in the SM~\cite{rad:sm},  to the 
processes under consideration show that the
total rates receive a correction less than 3\% for a Higgs with mass
about 120 GeV, thus validating our choice of $a_V = 1 +\Delta a_V$. 
Since radiative corrections to the processes we consider have
been computed not only in the SM~\cite{rad:sm} but also for the
MSSM~\cite{rad:mssm}, assessing the effects of these on the rates and
asymmetries and hence on the sensitivity for the anomalous couplings
will be the next logical step of this analysis which has shown
efficacy of these variables to probe these couplings.

Furthermore, in our analysis, we have imposed simple cuts on the kinematic
variables which virtually eliminate the non-Higgs backgrounds to the
particular final states under consideration. Since the latter involve
$b$ jets, we fold our results with realistic $b$-tagging efficiencies.
In addition, certain cuts also serve to enhance/suppress particular
contributions to the signal.  
For example, the additional cuts $R1$ and
$R2$  were introduced to enhance the contribution from $s-$channel and
$t-$channel diagrams respectively. If we look
at the observables pertaining to the $ZZH$ vertex, 
then all the couplings, except for $\Re(\tilde b_Z)$, are best
constrained with $R1$ cut, i.e, via the Bjorken diagram when $Z$ boson is
produced on-shell. The contributions from $\Im(b_Z)$ and $\Im(\tilde
b_Z)$ are proportional to $(r_e^2-l_e^2)$, and hence are small away from
the $Z$-pole. For $\Re(\tilde b_Z)$, however, the contribution from the
$t$-channel is roughly proportional to $(r_e^4+l_e^4)$ hence an analysis
with the $R2$ cut provides a better limit.

The situation is more complicated for the $WWH$ couplings. While the relevant 
final state, namely $\nu\bar\nu H$, receives large
contributions from $\tilde T$-odd coupling, the impracticality of 
measuring the momentum of an individual neutrino prevents us from 
isolating such contributions. Consequently, we are left with just two 
observables, which, coupled with $R1^\prime$ and $R2^\prime$ cuts, provide
probes of the $\tilde T$-even couplings. The bounds on the 
$\tilde T$-odd couplings are indirect and hence suffer from 
reduced sensitivity.
Furthermore, the contributions from non-zero anomalous $ZZH$ 
couplings cannot be eliminated in their entirety 
and are treated as contaminations in the determination of the $WWH$
couplings. Together, these two factors result in the bounds on the 
two $\tilde T$-odd $WWH$ couplings to be as weak as order unity. 
Note though that our entire formalism presupposes that 
the anomalous couplings are small and thus these limits on 
$\Re(\tilde b_W)$ and $\Im(b_W)$ are of little value. 

It is instructive to compare the results of our analysis with those 
of earlier
investigations. Ref.~\cite{t.han} had analyzed the case
of $ZZH$ couplings with (out) initial beam polarization. We find that
the effect of realistic cuts on kinematic variables required to
isolate the signal with the dominant final state with $b\bar b$ as
well as the finite $b$-tagging efficiency, reduces the possible limits
on $\tilde b_Z$ by about a factor of 2, compared to the ones quoted in
Ref.~\cite{t.han} with unpolarized beams. Needless to say, if the
reduction in rates implied by these cuts is neglected, our analysis
does reproduce the results of Ref.~\cite{t.han}.

In Ref.~\cite{Hagiwara:2000tk}, an optimal observable
analysis~\cite{oot} is performed, including along with an additional
anomalous $Z \gamma H$ coupling.  While such optimal variable analyses
generally indicate the maximum achievable sensitivity, the observables
constructed very often remain a little opaque with respect to the
physics they probe. The parameterization of
Ref.~\cite{Hagiwara:2000tk} is quite different from ours.  Still,
making use of the correlation matrices given by them, and putting the
$Z \gamma H$ coupling to zero, one may extract the limits their
analysis will imply for our parameterization. Doing this, we find
that, for the $\tilde T$-even couplings, our limits compare quite well
with those obtained in the analysis of Ref.~\cite{t.han}, implying
thereby that our simple $\tilde T$ observables indeed catch the
physics content of their optimal observables in this case. For the
$\tilde T$-odd couplings, our use of simple observable like the
expectation value of sign of ${\cal C}_{1,2}$ rather than the expectation 
value of the momentum correlator, $\langle{\cal C}_{1,2}\rangle$ causes a loss 
in sensitivity only by
a factor 4.  Given the fact that some of this is attributable to 
our use of realistic kinematic cuts,  $b$-tagging efficiencies etc, 
this is a very modest price to pay for the simplicity of the
observable.  The optimal observable analysis shows that the use of
$Z\to b\bar b$ and $Z\to \tau^+ \tau^-$ final states and polarization
of the beams can improve the sensitivity significantly. This is a very
good motivation for constructing analogous simple observables similar to the
ones constructed here.

To summarize, we have looked at the Higgs production processes at an 
$e^+e^-$ collider involving
$VVH$ coupling. We constructed several observables with appropriate $CP$ and
$\tilde T$ properties to probe various anomalous couplings incorporating
realistic cuts and detection efficiencies. 

Using these observables in the context of $ZZ$ fusion and
Higgstrahlung processes, we obtain stringent but realistic bounds on
the various anomalous $ZZH$ couplings, even while allowing for maximal
cancellations between the various individual contributions.  As for
the $WWH$ couplings, their effects cannot be fully isolated from those
of the $ZZH$ couplings. Nonetheless, we are able to derive quite
stringent bounds for the $\tilde T$-even subset even while accounting
for maximal contamination from the$ZZH$ sector. On the other hand, the
lack of suitable $\tilde T$-odd observables render the limits on the
$\tilde T$-odd $WWH$ couplings to be only indirect and thus poor.  We
reemphasize that all our asymmetries are simple to construct, have
specific $CP$ and $\tilde T$ properties to probe specific anomalous
coupling, and are robust against both the radiative correction to the
rates as well as systematic errors.

\section*{Acknowledgments}
We thank B. Mukhopadhyaya for collaboration in the initial phase of this
project and S. D. Rindani, M. Schumacher and K. Moenig for useful discussions
and S. K. Rai for reading the manuscript.
We would like to acknowledge support from the Department of Science and
Technology under project number SR/FIST/PSI-022/200, to the Center for High
Energy Physics, IISc, for the cluster which was used for computations. The 
work was also partially supported under the DST project 
SP/S2/K-01/2000-II and the Indo-French project IFC/3004-B/2004.
RKS wishes to thank Harish-Chandra Research Institute for
hospitality where this work was started and Council for Scientific and Industrial 
Research for the financial support.
DC thanks the DST, India for financial
assistance under the Swarnajayanti grant.

\appendix
\section{Partial cross-sections and anomalous couplings}
\label{pcs}
As mentioned in Section IIB, the total cross-section for $e^+e^- \to f\bar f
H$ receives contributions only from $CP$-even and $\tilde T$-even couplings
$\Delta a_V$ and $\Re(b_V)$, while the other couplings contribute to partial
cross-section in such a way that their net contribution to the total rate is
zero. In our analysis so far, we considered 
appropriately chosen partial cross-sections and
combined them to construct asymmetries. In its stead, we could, in principle,
have considered just the partial cross sections themselves and 
investigated their resolving power. This, we attempt now.

\begin{figure}[!h]
\epsfig{file=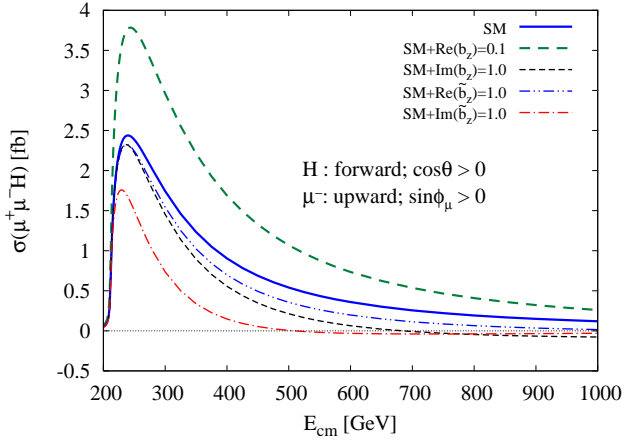,width=8.5cm,height=6cm}
\caption{\label{fig:fu-s} Partial cross-sections as a function of
c.m. energy for the process $e^+e^-\to\mu^+\mu^-H$ with Higgs boson in
the forward direction and the final state $\mu^-$ above the Higgs
boson's production plane.  A Higgs boson of mass 120 GeV is assumed.}
\end{figure}

To start with, we look
at $e^+e^- \to\mu^-\mu^+ H$ and constrain the Higgs boson to be 
in the forward direction
($\cos\theta_{e^-H}>0$) and $\mu^-$ to be above the Higgs production plane
($\sin\phi_\mu>0$). This partial cross-section, called
 ``forward-up'' and denoted
by $(F'U)$ in section III F, is plotted as a function of $E_{cm}$ in 
Fig.~\ref{fig:fu-s} for the SM. Also shown are the corresponding 
cross-sections when only one anomalous  coupling is non-zero. 
The large values of the anomalous couplings have been chosen 
to highlight the differences.
\begin{figure}[!h]
\epsfig{file=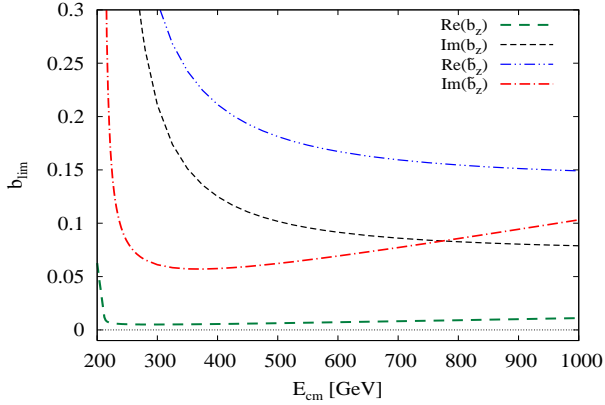,width=8.5cm,height=5.5cm}
\caption{\label{fig:lim-s}Limits on various non-standard coupling obtained using
Eq.(\ref{blim}) with $d=1$ and $\epsilon=0.01$ for an integrated luminosity of 
500 fb$^{-1}$ and using cross-sections shown in Fig.~\ref{fig:fu-s}. }
\end{figure}

We see that all four anomalous couplings contribute
to the $(F'U)$ partial cross-section, which can now be parametrized as
\begin{equation}
\sigma_{F'U} = \sigma^0 + \Re(b_Z)\sigma^1 + \Im(b_Z)\sigma^2 + \Re(\tilde b_Z)
\sigma^3 + \Im(\tilde b_Z) \sigma^4.
\label{sig:fu}
\end{equation}
If only one anomalous coupling ${\cal B}_i$ were to be non-zero, then the
measurement of this partial cross-section would be sensitive to  
\begin{equation}
 |{\cal B}_i| \geq \frac{d}{|\sigma^i|} \sqrt{\frac{\sigma^0}{\cal L} +
\epsilon^2(\sigma^0)^2}.
\label{blim}
\end{equation}
Here $d$ is the degree of statistical significance, ${\cal L}$ is the
integrated luminosity of the $e^+e^-$ collider and $\epsilon$ is the
fractional systematic error.  In Fig.~\ref{fig:lim-s}, we show a
simple ($d=1,\epsilon=0.01$) limit on anomalous couplings, obtained
using Fig.~\ref{fig:fu-s}, for an integrated luminosity of 500
fb$^{-1}$.  A measurement of $\sigma_{F'U}$ will thus be sensitive to
values of ${\cal B}_i$ lying above the corresponding curve.

\begin{figure}[!ht]
\epsfig{file=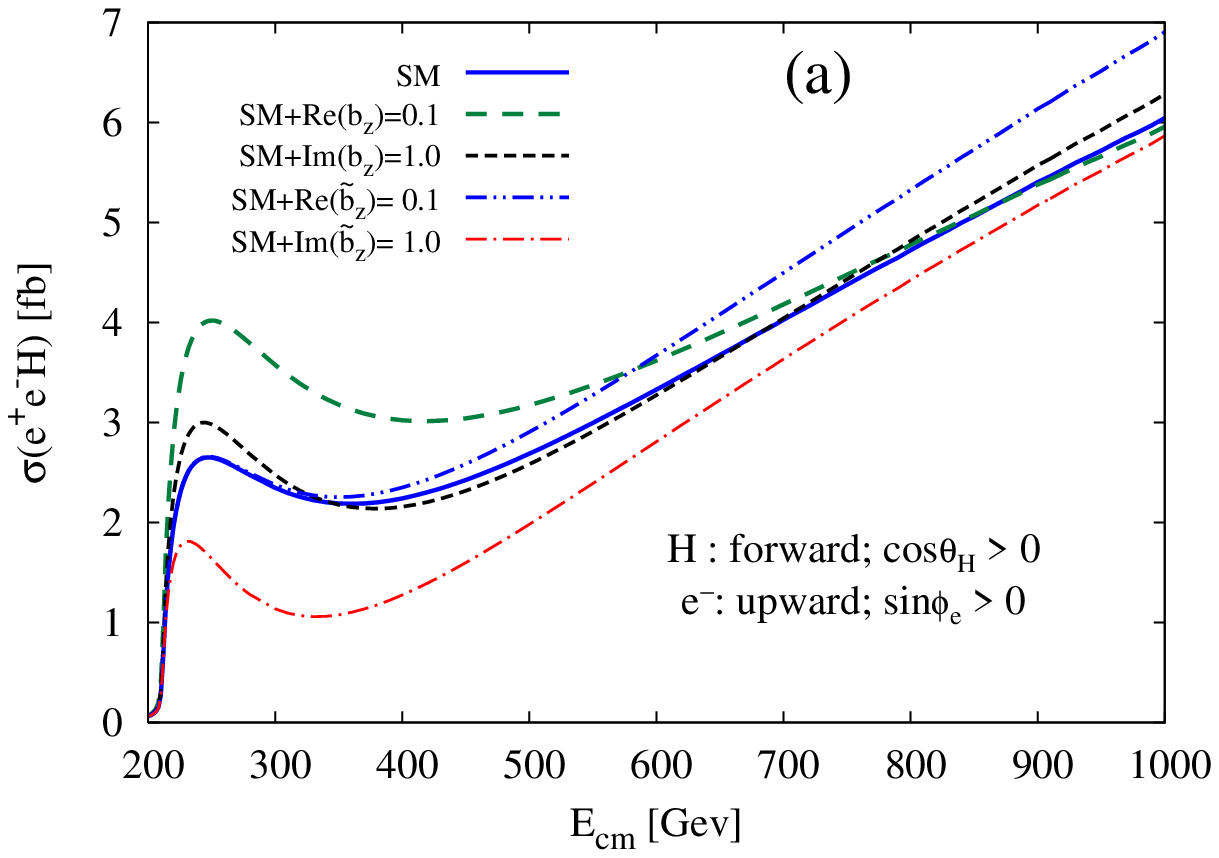,width=8.5cm,height=6.5cm}
\epsfig{file=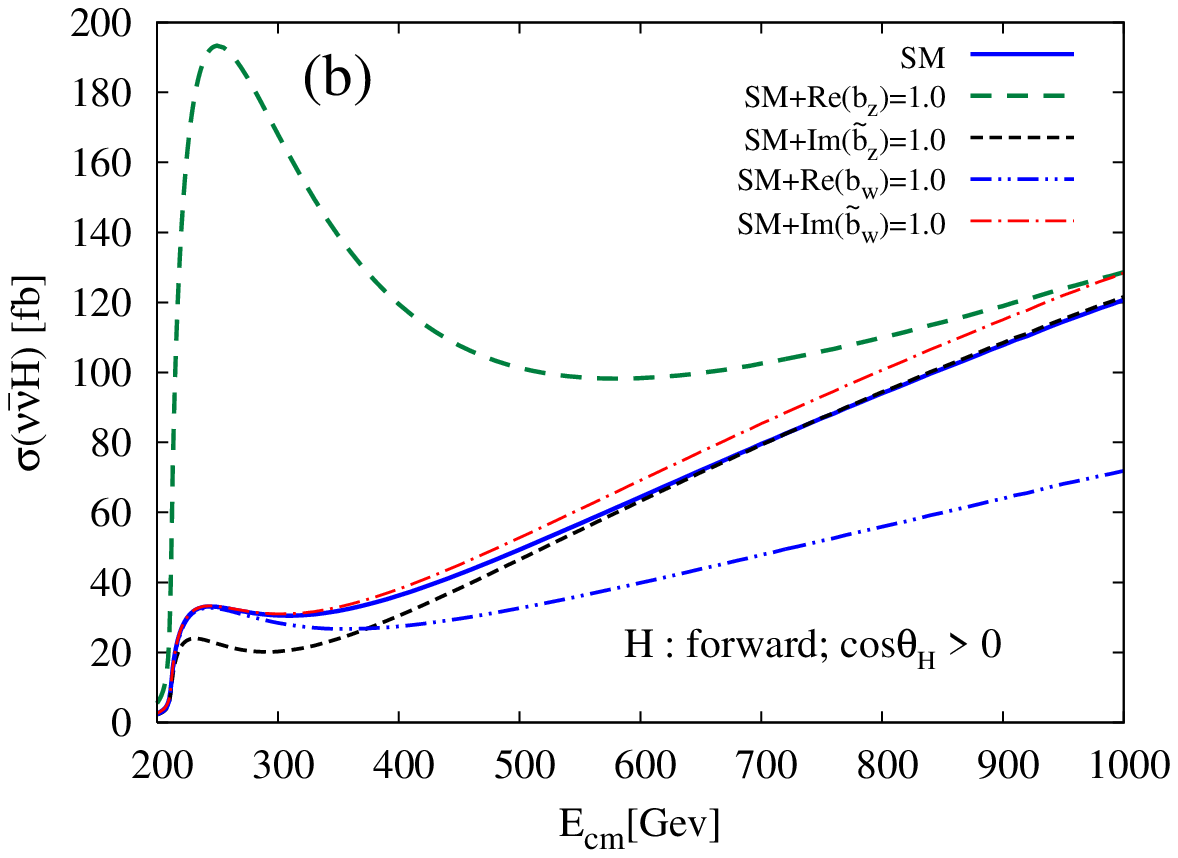,width=8.5cm,height=6.5cm}
\caption{\label{fig:hnn} Cross-sections as functions of the c.m. energy
for a Higgs boson of mass 120 GeV and particular values for the
anomalous couplings. 
(a) Partial cross-sections for $e^+e^-\to e^+e^-H$ with forward Higgs
boson and the final state $e^-$ above the Higgs boson's production
plane.
(b) Partial cross-sections for
$e^+e^-\to\nu\bar\nu H$ with the Higgs boson in the forward direction.
}
\end{figure}

A similar exercise can be done 
for $e^+e^-\to e^+e^-H$, and in Fig.~\ref{fig:hnn}(a) we display 
the $(F'U)$ partial cross-section for the same as a function of
the center-of-mass energy. 
The presence of an additional
$t$-channel diagram changes the $E_{cm}$ behaviour of the partial
cross-section and hence that of the limits that could be inferred 
in a fashion analogous to  Fig.~\ref{fig:fu-s}.

For $\Re(\tilde b_Z)$ and $\Im(\tilde b_Z)$, the 
$3\sigma$ bounds of 
Table~\ref{tab:ind-lim} are much better than the $1\sigma$ limits shown in
Fig.~\ref{fig:lim-s}. This indicates
that asymmetries with appropriate symmetry properties and combinations of
various final states can be used efficiently to obtain stringent constraints on
anomalous couplings. 

For  $\Im(\tilde b_Z)$, on the other hand, the limit obtained
using $A^{com}_\mu$ is only comparable to the one
obtained using just the partial rate $\sigma(F'U)$ after accounting for 
degrees of significance. However, the limit from
$\sigma(F'U)$ is subject to the assumption that all other anomalous
couplings are zero, while the one obtained using $A^{com}_\mu$ is
independent of any other anomalous coupling. Once again this 
underscores the
importance of specific observables, such as $A^{com}$, which receive a
contribution from only one of the anomalous coupling, thus allowing
us to obtain a robust constraint.

And finally, we look at the 
$e^+e^-\to \nu_e\bar\nu_e H$ channel, where both $WWH$ and $ZZH$ vertices
contribute thus doubling the number of anomalous couplings
involved. Since the final state fermions, the neutrinos, are not
detectable, it is meaningless to construct the partial cross-section
$(F'U)$. Instead we add $(F'U)$ and $(F'D)$ to form the ``forward''
cross-section, {\em i.e.} the Higgs boson is constrained to be in the
forward direction, and the $E_{cm}$ dependence is displayed in
Fig.~\ref{fig:hnn}(b). The size of the anomalous contribution in these
figures gives an idea about the sensitivity to that particular
coupling.

\section{Expressions for $|M^2|$ }
\label{M2}
In this appendix, we list the square of invariant matrix element
for the various processes considered in the text. To begin with,
we define the fermion-$Z$ vertices by
\[
\frac{i g}{\sqrt{2}} \, \gamma_\mu \,
        (\ell_e \, P_L + r_e \, P_R) \ ,
\]
In considering a process such as
$e^-(p_1) e^+(p_2) \to f(p_3) \bar f(p_4)  h(p_5)$, it is
further convenient to devise a notation for scalar
products such as
\begin{eqnarray}
\pdot{i}{j}  \equiv  p_i \cdot p_j, \  
{\cal A}  \equiv  \epsilon_{\mu \nu \sigma \rho} \, p_1^\mu \, p_2^\nu
\, p_3^\sigma \, p_4^\rho
\end{eqnarray}
and similarly for the multitude of propagators that
one encounters, namely
\begin{eqnarray}
{\cal S}_{ij} & \equiv &
  \left( (p_i+p_j)^2 - m_Z^2 + i \Gamma_Z \, m_Z \right)^{-1}\\
{\cal Z}_{ij} & \equiv &
  \left( (p_i-p_j)^2 - m_Z^2 + i \Gamma_Z \, m_Z \right)^{-1}\\
  {\cal W}_{ij} & \equiv &
  \left((p_i-p_j)^2 - m_W^2 + i \Gamma_W \, m_W \right)^{-1}
\end{eqnarray}
we have, for where $f (\neq e, \nu_e)$ is any massless fermion,
\begin{widetext}
\begin{equation}
\begin{array}{rcl}
|{\cal M}|^2 & = g^4 \ \left| {\cal S}_{12} \, {\cal S}_{34} \right|^2 \; 
   & \displaystyle
\Bigg[
|a_Z|^2 \left\{ (\lesq \, \lqsq +  \resq \, \rqsq) \,  \pdot14 \, \pdot23
 +  (\lesq \, \rqsq  +  \resq \, \lqsq) \,  \pdot13 \, \pdot24 \right\}
\\[2ex]
&& \displaystyle
 -  \frac{\reaZbZconj}{m_Z^2}\,  \Bigg( (\lesq \, \lqsq +\resq \, \rqsq ) 
\, (\pdot14 +  \pdot23 ) \, \left\{ \pdot13 \, \pdot24  - \pdot12 \, \pdot34 \,
 - \pdot14 \, \pdot23 \, \right\}\\
&&  \displaystyle \hspace*{6em}
+ (\lesq \, \rqsq  + \resq \, \lqsq )\, 
 (\pdot13 + \pdot24) \,
\left\{ \pdot14 \, \pdot23   - \pdot12 \, \pdot34
 - \pdot13 \, \pdot24  \right\} \Bigg)\\[3ex]
& & \displaystyle - \, \frac{\imaZbZconj}{m_Z^2}\,  \, {\cal A}
 \Bigg((\lesq \, \lqsq - \resq \, \rqsq) \, ( \pdot14 - \pdot23 ) 
+  (\resq \, \lqsq - \lesq \, \rqsq)\,  ( \pdot13 - \pdot24 )  \Bigg)
\\[3ex]
& & \displaystyle + \frac{\reaZbtilZconj}{m_Z^2} \, {\cal A}  \, \Bigg( 
(\lesq \, \lqsq  +\resq \, \rqsq )\,( \pdot14 + \pdot23 ) 
 - (\lesq \, \rqsq  + \resq \, \lqsq )\,  ( \pdot13 + \pdot24 )  \Bigg)
\\[3ex]
& & \displaystyle - \; 
\frac{\imaZbtilZconj}{m_Z^2} \, \Bigg( (\lesq \, \lqsq -\resq \, \rqsq)
\, (\pdot23 - \pdot14 ) 
\left\{ \pdot12 \, \pdot34 - \pdot13 \, \pdot24 + \pdot14 \, \pdot23  \right\} 
\\
&& \displaystyle \hspace*{6em} + (\resq \, \lqsq - \lesq \, \rqsq)  
(\pdot13 - \pdot24)  \, \left\{  \pdot12 \,\pdot34 - \pdot14 \, \pdot23
 + \pdot13 \, \pdot24     \right\}  \Bigg)\Bigg] 
\end{array}
    \label{mesq_quark}
\end{equation}
For $f = e$, the expressions are a bit more complicated, and
\begin{eqnarray}
|{\cal M}|^2 & = & g^4 \;
\Bigg[{\cal R}_S \; \left| {\cal S}_{12} \, {\cal S}_{34} \right|^2
 + {\cal R}_T \; \left| {\cal Z}_{13} \, {\cal Z}_{24} \right|^2 \;
 - 2 \, Re( {\cal R}_I \,
 {\cal S}_{12}^* \, {\cal S}_{34}^*  {\cal Z}_{13} \, {\cal Z}_{24} )
 \Bigg] 
\end{eqnarray}
where,
\begin{equation}
\begin{array}{rcl}
{\cal R}_S & = & \displaystyle 
|a_Z|^2 \left[ 2 \lesq \resq   \, \pdot13 \, \pdot24
       + (\lefour + \refour)  \, \pdot14 \, \pdot23     \right] 
- \, \frac{\imaZbZconj}{m_Z^2} \, (\lefour - \refour) \,
 {\cal A}   ( \pdot14 - \, \pdot23 ) 
\\
& + & \displaystyle 
\frac{\reaZbZconj}{m_Z^2} \; \Bigg[(\lefour  + \refour) \, (\pdot14 + \pdot23) \
 \left\{  \pdot12 \, \pdot34  - \, \pdot13 \, \pdot24 \
             + \, \pdot14  \, \pdot23 \right\} 
 \\
&&\displaystyle \hspace*{3em}
+ 2 \, \lesq \resq   \, (\pdot13  + \pdot24 ) \,
  \left\{ \pdot12 \, \pdot34 + \pdot13 \, \pdot24 - \pdot14 \, \pdot23
  \right\} \Bigg] \\[2ex]
&+ & \displaystyle \frac{\imaZbtilZconj}{m_Z^2} \, (\lefour - \refour) \, (\pdot14 - \pdot23) \,
  \left\{  \pdot12 \, \pdot34  - \, \pdot13  \, \pdot24 + \pdot14 \,
  \pdot23 \right\} \\[2ex]
&+& \displaystyle
\frac{\reaZbtilZconj}{m_Z^2} \, {\cal A} \, \left[(\lefour + \refour) \,  
( \pdot14 + \, \pdot23 ) - 2 \,  \lesq \resq   ( \pdot13 + \pdot24 ) \right], 
\end{array}
\end{equation}
\begin{equation}
\begin{array}{rcl}
{\cal R}_T & = & \displaystyle
|a_Z|^2 \, \left[ (\lefour + \refour) \, \pdot14 \, \pdot23
+ 2 \, \lesq \, \resq  \, \pdot12 \, \pdot34 \right]
- \; 
\frac{\imaZbZconj}{m_Z^2} \; (\refour - \lefour) \,  {\cal A}
         \, ( \pdot14 -  \pdot23 ) \\[2ex]
& + & \displaystyle
\frac{\reaZbZconj}{m_Z^2} \; \Bigg[ (\lefour + \refour) \, (\pdot14 + 
\pdot23 ) \, \left\{  - \, \pdot12 \, \pdot34 + \pdot13 \, \pdot24
+ \pdot14 \, \pdot23 \right\} \nonumber \\
&& +  2 \, \lesq \resq  \, (\pdot12 \, + \pdot34 ) \; \left\{   - \pdot13 \, 
\pdot24  -  \pdot12 \, \pdot34 + \pdot14 \, \pdot23  \right\}  \Bigg]
\nonumber\\
&+ & \displaystyle\frac{\imaZbtilZconj}{m_Z^2} \; (\refour - \lefour) \; (\pdot23 - \pdot14) 
\; \left\{ - \pdot12 \, \pdot34 + \, \pdot13 \, \pdot24 + \pdot14 \, \pdot23 
\, \right\}\nonumber\\
&- & \displaystyle\frac{\reaZbtilZconj}{m_Z^2} \;  {\cal A} \; \left[
 (\lefour + \refour) (  \pdot14 + \pdot23 ) + 2 \,  \lesq \, \resq \, 
 ( \pdot12 + \pdot34 ) \right], 
\end{array}
\end{equation}
\begin{equation}
\begin{array}{rcl}
{\cal R}_I & = & 
  \displaystyle -  |a_Z|^2 \, ( \lefour + \refour) \; \pdot14 \, \pdot23
\\[2ex]
& + & \displaystyle 
\frac{(\lefour - \refour)}{m_Z^2} \;  ( \pdot23  -  \pdot14 ) 
\; \left[ i \, \reaZbZconj \;  {\cal A} 
+ i \, \reaZbtilZconj \, 
( \pdot13 \, \pdot24 - \pdot12 \, \pdot34)
+ \imaZbtilZconj
 \, \pdot14 \, \pdot23 
\right]
\\[3ex]
&-  & \displaystyle 
\frac{(\lefour  + \refour)}{m_Z^2}  \; ( \pdot14 + \pdot23 ) 
\; \left[ i \, \imaZbZconj  
             \; ( \pdot12 \, \pdot34  - \pdot13 \, \pdot24)  
+ i \, \imaZbtilZconj \, {\cal A} 
+  \reaZbZconj \, \pdot14  \, \pdot23
\right]
\end{array}
\end{equation}
And finally, for $f = \nu_e$,
\begin{eqnarray}
|{\cal M}|^2 & = &
g^4\Bigg[\ell_\nu^2{\cal R}_S \; \left| {\cal S}_{12} \, {\cal S}_{34} \right|^2
+ \; {\cal R}_T \; \left| {\cal W}_{13} \, {\cal W}_{24} \right|^2
 - \ell_\nu \, \ell_e \Re( {\cal R}_I \, {\cal S}_{12}^* \, 
 {\cal S}_{34}^*  {\cal W}_{13} \,  {\cal W}_{24})  \Bigg] 
\end{eqnarray}
where,
\begin{eqnarray}
{\cal R}_S & = & |a_Z|^2 \; \left[\lesq \, \pdot14 \, \pdot23 + \resq \, \pdot13 
\, \pdot24 \right] 
+ \frac{\imaZbZconj}{m_Z^2} \;
 {\cal A} \; \Bigg[\lesq \,  ( \pdot23 - \pdot14  ) +  \resq \,  ( \pdot24 - 
 \pdot13 ) \Bigg] \nonumber\\
& + & \frac{\reaZbZconj}{m_Z^2} \, \Bigg[ \lesq \,  (\pdot14 +  
\pdot23 ) \,    \left\{ \pdot12  \, \pdot34  - \, \pdot13 \, \pdot24
 + \pdot14 \, \pdot23  \right\}	
+ \resq \,(\pdot13 +  \pdot24) \,
  \left\{ \pdot12  \, \pdot34  +  \pdot13 \, \pdot24
          -  \pdot14 \, \pdot23 \right\}  \Bigg] \nonumber\\
& + & \frac{\imaZbtilZconj}{m_Z^2} \, \Bigg[\lesq \, (\pdot14 -  \pdot23 ) \,
 \left\{  \pdot12 \, \pdot34 - \, \pdot13 \, \pdot24 \, + \pdot14 \, \pdot23 
 \right\} +  \resq \,  (\pdot24 - \pdot13 ) \,
 \left\{   \pdot12  \, \pdot34  + \pdot13 \, \pdot24 - \pdot14 \, \pdot23
 \right\} \Bigg] \nonumber\\
& - & \frac{\reaZbtilZconj}{m_Z^2} \; {\cal A} \; \Bigg[\resq \,(\pdot13 + 
\pdot24 ) - \lesq \,  ( \pdot14 + \, \pdot23 ) \Bigg],
\end{eqnarray}
\begin{eqnarray}
{\cal R}_T & = &|a_W|^2 \, \pdot14 \, \pdot23 +\frac{\reaWbWconj}{m_W^2}  \,
(\pdot14 + \pdot23 ) \, \left\{  - \, \pdot12  \, \pdot34 + \, \pdot13  \, 
\pdot24 + \, \pdot14 \, \pdot23 \right\} + \frac{\imaWbWconj}{m_W^2} \; {\cal A}
\; (\pdot14 - \pdot23 ) \nonumber\\
& + & \frac{\imaWbtilWconj}{m_W^2} \; (\pdot14 - \, \pdot23 ) \; \left\{
  - \, \pdot12 \, \pdot34  + \, \pdot13 \,  \pdot24  + \, \pdot14 \, \pdot23 
  \right\} - \frac{\reaWbtilWconj}{m_W^2} \; {\cal A}  \; ( \pdot14 + \pdot23 )
\end{eqnarray}
\begin{eqnarray}
{\cal R}_I &=& - 2 \, a_W \, a_Z^*    \, \pdot14 \, \pdot23 - \frac{a_W \,
b_Z^*}{m_Z^2} \, \Bigg[ (\pdot14 + \pdot23  ) \, \left\{   \pdot12 \, \pdot34
- \, \pdot13\, \pdot24 + \pdot14 \pdot23\right\}
+  i \; {\cal A}  \;( \pdot14  - \pdot23 ) \Bigg] \nonumber\\
& - & i \; \frac{a_W \, \btilZ^*}{m_Z^2} \, \Bigg[ (\pdot23 - \pdot14 ) \,
\left\{ \pdot12 \, \pdot34    - \pdot13 \, \pdot24 +  \pdot14 \, \pdot23 
\right\} - i \; {\cal A} \; (  \pdot14 + \pdot23 )  \Bigg]\nonumber\\
& + &  \frac{a_Z^* \, b_W}{m_W^2} \, \Bigg[ (\pdot14 + \pdot23 ) \,
\left\{   \pdot12 \, \pdot34  - \, \pdot13 \, \pdot24 - \, \pdot14 \, 
\pdot23 \right\}  - i  \;  {\cal A} \; ( \pdot14  - \pdot23  )
\Bigg]\nonumber\\
& + &  i \; \frac{a_Z^* \, \btilW}{m_W^2} \, \Bigg[ ( \pdot23 - \pdot14 ) \,
\left\{ - \pdot12 \, \pdot34 +  \pdot13 \, \pdot24 + \pdot14 \, \pdot23 
\right\} - i \; {\cal A}    \;  ( \pdot14 + \pdot23 ) \Bigg]
\end{eqnarray}
In the propagators, we ignore the contribution proportional to $\Gamma_V$
except for ${\cal S}_{34}$, which goes on-shell, and cannot be ignored, in
general.
\end{widetext}

\end{document}